\documentclass[a4paper,11pt]{article}
\usepackage{pos,bm}

\newcommand{\pvec}{\bm{p}}
\newcommand{\beq}{\begin{equation}}
\newcommand{\eeq}{\end{equation}}
\newcommand{\beqs}{\begin{eqnarray}}
\newcommand{\eeqs}{\end{eqnarray}}
\newcommand{\bitem}{\begin{itemize}\item}
\newcommand{\eitem}{\end{itemize}}

\newcommand{\ket}[1]{\vert #1\rangle}

\newcommand{\me}[3]{\langle #1\vert\ #2\ \vert #3\rangle}

\newcommand{\wvec}{\mathbf{w}}

\newcommand{\xvec}{\mathbf{x}}

\newcommand{\qqa}{\bm{q}^2_{{\rm cm},a}}

\newcommand{\dvec}{\bm{d}}
\newcommand{\Pvec}{\bm{P}}
\newcommand{\svec}{\bm{s}}
\newcommand{\nn}{\nonumber}

\newcommand{\Ecm}{E_{\rm cm}}
\newcommand{\nvec}{\bm{n}}
\newcommand{\zvec}{\bm{z}}

\title{Nucleon scattering from lattice QCD}

\author*[a]{Colin Morningstar}

\affiliation[a]{Department of Physics, Carnegie Mellon University, 
Pittsburgh, Pennsylvania 15213, USA}

\emailAdd{cmorning@andrew.cmu.edu}

\abstract{Recent results from lattice QCD on baryon resonances and
meson-baryon, baryon-baryon scattering are presented. Such scattering processes
and resonances can be determined in lattice QCD by first obtaining the finite-volume
energy spectrum of stationary states involving meson-baryon and baryon-baryon
systems.  A well-known quantization condition involving the scattering $K$-matrix
and a complicated ``box matrix'' also yields a finite-volume energy spectrum.
By appropriately parametrizing the scattering $K$-matrix, the best fit values
of the $K$-matrix parameters are those which produce a finite-volume spectrum
which best matches that obtained from lattice QCD.  The $\Delta$ resonance, a 
recent study of the two-pole nature of scattering near the $\Lambda(1405)$, and 
$NN$ scattering in the $SU(3)$ flavor limit are highlighted.}

\FullConference{The 11th International Workshop on Chiral Dynamics (CD2024)\\
 26-30 August 2024\\ Ruhr University Bochum, Germany}

\begin{document}
\maketitle

\section{Introduction}
In recent years, lattice QCD studies have finally been able to compute
the masses and decay widths of unstable hadron resonances, such as the $\rho$
or $\Lambda(1405)$ resonances.  Such studies require first 
computing the finite-volume energies of the multi-hadron states into which the 
resonances decay using Markov-chain Monte Carlo path integration.  Next,
functional forms of the scattering amplitudes are introduced which involve various 
parameters, and these forms are inserted into a well-known quantization condition
involving the scattering $K$-matrix and a complicated ``box matrix''
which yields a finite-volume spectrum dependent on the scattering amplitude
parameters.  Finally, the best-fit values of these parameters are found as those
which yield a finite-volume energy spectrum which best matches that obtained from 
the lattice QCD calculations.  The resonance properties follow from the
scattering amplitudes.

An important step in such computations is the evaluation of the energies of the
stationary states in finite-volume involving multi-hadron contributions. These 
energies are extracted from Monte Carlo estimates of temporal correlations 
involving judiciously-designed quantum field operators that create the
needed states of interest.  To evaluate such correlators, quark propagators 
from a variety of source sites on the lattice must be contracted together.  The 
quark propagators themselves are the inverses of an exceedingly large matrix, 
but fortunately, only the matrix products of the inverse with the various 
source sites is needed.  For correlators involving only single-hadron operators,
translational invariance can be used to limit the number of source sites to 
a small few, but for correlators involving multi-hadron operators, all
spatial sites on a source time slice must be used.  For this reason, reliable 
estimates of such correlations involving multi-hadron operators were not 
feasible to attain until recently. Novel techniques, such as a quark-field 
smearing scheme known as Laplacian Heaviside (LapH), have now made such 
reliable estimates feasible. LapH quark smearing projects the quark propagators
into a smaller subspace spanned by various eigenvectors of the gauge-covariant
Laplacian, allowing the use of all spatial sites on a source time slice in a 
feasible manner. 

Recent results from lattice QCD on baryon resonances and meson-baryon, 
baryon-baryon scattering are presented in this talk.  After outlining how
such studies are accomplished, a recent investigation of the $\Delta$ resonance,
a  recent study of the two-pole nature of scattering near the $\Lambda(1405)$, and 
$NN$ scattering in the $SU(3)$ flavor limit are highlighted.

\section{Outline of Methodology}

The first step in lattice QCD studies of resonance and scattering properties
is to evaluate the finite-volume energies of stationary states corresponding to
the relevant decay products for variety of total momenta and symmetry 
representations.  Stationary-state energies are extracted from an  $N\times N$ 
Hermitian temporal correlation matrix 
   $C_{ij}(t)
   = \langle 0\vert\, O_i(t\!+\!t_0)\, \overline{O}_j(t_0)\ \vert 0\rangle
   $
involving carefully designed operators $O_j(t)=O_j[\overline{\psi},\psi,U]$ 
comprised of quark $\psi,\overline{\psi}$ and gluon $U$ field operators which create
the states of interest. The temporal correlators are obtained from path integrals over
the fields
\beq
  C_{ij}(t)= \frac{ \int {\cal D}(\overline{\psi},\psi,U)\ \ 
   O_i(t+t_0)\ \overline{O}_j(t_0)\ \ \exp\left(-S[\overline{\psi},\psi,U]\right)}{  
  \int {\cal D}(\overline{\psi},\psi,U)
 \ \exp\left(-S[\overline{\psi},\psi,U]\right)},
\eeq
where the action in imaginary time has the form
\beq
  S[\overline{\psi},\psi,U] = \overline{\psi}\ K[U]\ \psi + S_G[U],
\eeq
and where $K[U]$ is the fermion Dirac matrix and $S_G[U]$ is the gluon action.
The integrals over the Grassmann-valued quark/antiquark fields can be done exactly,
leaving expressions of the form
  \beq
  C_{ij}(t)= \frac{ \int {\cal D}U\ \det K[U]\ 
  \left( K^{-1}[U]\cdots K^{-1}[U] +\dots\right)\ \ \exp\left(-S_G[U]\right)}{  
  \int {\cal D}U\ \det K[U]
 \ \exp\left(-S_G[U]\right)}.
  \eeq
For the integrations over the gluon fields, we must resort to the Monte Carlo method,
which requires formulating QCD on a space-time lattice (usually hypercubic),
with quark fields residing on the sites and the gluon field residing on the links 
between lattice sites.  A Markov chain is used to generate a sequence of gauge-field
configurations $ U_1, U_2,\dots, U_N$ using the Metropolis method\cite{Metropolis:1953am}
with a complicated global updating proposal, such as RHMC\cite{Clark:2006fx}, 
which solves Hamilton equations with Gaussian
momenta.  The $\det K$ is usually estimated by an integral over pseudo-fermion fields.
The correlators are then estimated using the ensemble of gauge configurations
generated by the above procedure.  Systematic errors include discretization and
finite volume effects.  To speed up computations, unphysically large quark masses
are often used.

\begin{figure}
 \begin{center}
 \includegraphics[width=3.5in]{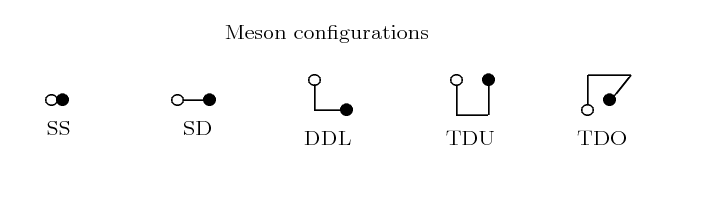}\\[-20pt]
 \includegraphics[width=4.0in]{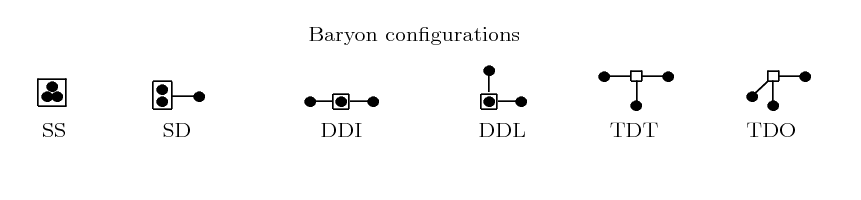}
 \end{center}
\vspace*{-8mm}
\caption{The spatial arrangements of the quark-antiquark meson operators (top)
and the three-quark baryon operators (bottom) that we use. Smeared quarks fields
are shown as solid circles, each hollow 
circle indicates a smeared antiquark field, the solid line segments indicate 
covariant displacements, and each hollow box indicates the location of a 
Levi-Civita color coupling. 
\label{fig:shopers}}
\end{figure}

Single-hadron operators are constructed using covariantly-displaced LapH-smeared 
quark fields as building blocks.  Stout link smearing\cite{Morningstar:2003gk} 
is used for the gauge field links $\widetilde{U}_j(x)$,
as well as Laplacian-Heaviside (LapH)\cite{HadronSpectrum:2009krc,Morningstar:2011ka}
smeared quark fields
   \beq
    \widetilde{\psi}_{a\alpha}(x) =
      {\cal S}_{ab}(x,y)\ \psi_{b\alpha}(y),
     \qquad {\cal S} = 
     \Theta\left(\sigma_s^2+\widetilde{\Delta}\right),
   \eeq
where $\widetilde{\Delta}$ denotes a 3-dimensional gauge-covariant Laplacian 
defined in terms of the stout links $\widetilde{U}$.
Displaced quark fields are defined by
  \beq
 q^A_{a\alpha j}= D^{(j)}\widetilde{\psi}_{a\alpha}^{(A)},
 \qquad  \overline{q}^A_{a\alpha j}
 = \widetilde{\overline{\psi}}_{a\alpha}^{(A)}
  \gamma_4\, D^{(j)\dagger}
  \eeq
where the displacement $D^{(j)}$ is a product of smeared links
\beq
 D^{(j)}(x,x^\prime) =
 \widetilde{U}_{j_1}(x)\ \widetilde{U}_{j_2}(x\!+\!d_2)
 \ \widetilde{U}_{j_3}(x\!+\!d_3)\dots \widetilde{U}_{j_p}(x\!+\!d_p)
  \delta_{x^\prime,\ x+d_{p+1}}.
\eeq
In the above equations, $a,b$ are color indices, $\alpha$ is a Dirac spin
index, and $j$ labels a displacement path of gauge links in directions $j_1,
j_2,\cdots$.
A variety of displacements can be used to build up the needed orbital and
radial structure, as shown in Fig.~\ref{fig:shopers}.  So-called elemental
quark-antiquark and three-quark operators which create a definite momentum 
$\pvec$ are defined by
 \beqs
 \overline{\Phi}_{\alpha\beta}^{AB}(\pvec,t)&=&
 \textstyle\sum_{\bm{x}} e^{i\pvec\cdot(\xvec+\frac{1}{2}(\bm{d}_\alpha+\bm{d}_\beta))}
   \delta_{ab}\ \overline{q}^B_{b\beta}(\bm{x},t)\ q^A_{a\alpha}(\bm{x},t),
 \\
  \overline{\Phi}_{\alpha\beta\gamma}^{ABC}(\pvec,t)&=& 
 \textstyle\sum_{\bm{x}} e^{i\pvec\cdot\xvec}\varepsilon_{abc}
\ \overline{q}^C_{c\gamma}(\bm{x},t)
\ \overline{q}^B_{b\beta}(\bm{x},t)
\ \overline{q}^A_{a\alpha}(\bm{x},t).
\eeqs
In these operators, $A,B,C$ are quark flavor indices, $a,b,c$ are color indices,
$\bm{d}_\alpha, \bm{d}_\beta$ are the spatial displacements of the $q, \overline{q}$
fields, respectively, from site $\bm{x}$,
and $\alpha,\beta,\gamma$ denote compound indices incorporating both
a Dirac spin index and a displacement path.
Group theoretical projections onto the irreducible representations (irreps) 
of the lattice symmetry group are then employed to create the final
single meson and single baryon operators:
\beq
  \overline{M}_{l}(t)= c^{(l)\ast}_{
 \alpha\beta}\ \overline{\Phi}^{AB}_{\alpha\beta}(t)\qquad\qquad
  \overline{B}_{l}(t)= c^{(l)\ast}_{
 \alpha\beta\gamma}\ \overline{\Phi}^{ABC}_{\alpha\beta\gamma}(t).
\eeq
In the above, $\alpha,\beta,\gamma$ again indicate compound indices incorporating both
a Dirac spin index and a displacement path, and $l$ is the final index
which labels the hadron operator.

Stationary-state energies are extracted from the temporal correlations
via their spectral representation
   \beq
   C_{ij}(t) = \sum_n Z_i^{(n)} Z_j^{(n)\ast}\ e^{-E_n t},
   \qquad\quad Z_j^{(n)}=  \me{0}{O_j}{n},
   \eeq
which neglects small temporal wrap-around contributions, where the 
energies $E_n$ are discrete in finite volume. It is not practical to do 
fits using above form, so one way to proceed is to define a new correlation
matrix $\widetilde{C}(t)$ using a single rotation
   \beq
   \widetilde{C}(t) = U^\dagger\ C(\tau_0)^{-1/2}\ C(t)\ C(\tau_0)^{-1/2}\ U
   \eeq
where the columns of $U$ are the eigenvectors of
   $C(\tau_0)^{-1/2}\,C(\tau_D)\,C(\tau_0)^{-1/2}$.
One then chooses $\tau_0$ and $\tau_D$ large enough so $\widetilde{C}(t)$ 
remains diagonal for $t>\tau_D$ within statistical errors.
Two-exponential fits to the diagonal rotated correlators
$\widetilde{C}_{\alpha\alpha}(t)$ then yield the energies $E_\alpha$ and 
overlaps $Z_j^{(n)}$. Energy shifts from non-interacting values can also
be obtained from  single exponential fits to a suitable ratio of 
correlators, but such fits must be cautiously done in combination with
fits to correlators that are not ratios.

Once single hadron operators are designed, two- and three-hadron operators
are straightforward to produce as appropriate superpositions of products
of single-hadron operators of definite momenta
    \beq
    c^{I_{3a}I_{3b}}_{\pvec_a\lambda_a;\ \pvec_b\lambda_b}
     \ B^{I_aI_{3a}S_a}_{\pvec_a\Lambda_a\lambda_a i_a}
     \  B^{I_bI_{3b}S_b}_{\pvec_b\Lambda_b\lambda_b i_b}
    \eeq
for fixed total momentum $\pvec=\pvec_a+\pvec_b$ and
fixed $\Lambda_a, i_a, \Lambda_b, i_b$.  Group theory projections onto 
the little group of $\pvec$ and isospin irreps are then carried out.
It is crucial to know and fix all phases of the single-hadron operators 
for all momenta, and this is usually done by selecting a reference
momentum direction $\pvec_{\rm ref}$, then for each momentum $\pvec$, 
selecting one reference rotation $R_{\rm ref}^{\pvec}$ that 
transforms $\pvec_{\rm ref}$ into $\pvec$.  This method is efficient
for creating large numbers of multi-hadron operators.

The idea that the finite-volume energies obtained in lattice QCD can be related to the
infinite-volume scattering $S$-matrix was first discussed in  
Refs.~\cite{Luscher:1990ux,Luscher:1991cf}.  These calculations
were later revisited in Ref.~\cite{Rummukainen:1995vs,Kim:2005gf} in the case of a single channel 
of identical spinless particles, and subsequent works have generalized their results
to treat multi-channels with different particle masses and 
nonzero spins\cite{Briceno:2014oea}. 
Our methodology for calculating scattering phase shifts was presented in
Ref.~\cite{Morningstar:2017spu} and is summarized below.

Since it is easier to parametrize a real symmetric matrix than a unitary 
matrix, one usually employs the real and symmetric 
$K$-matrix\cite{Wigner:1946zz,Wigner:1947zz}, defined using the $S$-matrix by
\beq
   S = (1+iK)(1-iK)^{-1} = (1-iK)^{-1}(1+iK). 
\eeq
Rotational invariance implies that
\beq 
  \langle J'm_{J'}L^\prime S^\prime a'\vert\ K
\ \vert Jm_JLS  a\rangle = \delta_{J'J}\delta_{m_{J'}m_J}
 \ K^{(J)}_{L'S'a';\ LS a}(\Ecm).
\eeq
We use an orthonormal basis of states, each labelled by $\ket{Jm_JLS a}$, where $J$ is
the total angular momentum of the two particles in the center-of-momentum frame, $m_J$ is 
the projection of the total angular momentum onto the $z$-axis, $L$ is the orbital angular 
momentum of the two particles in the center-of-momentum frame (not to be confused with the
lattice length here), $S$ in the basis vector is 
the total spin of the two particles (not the scattering matrix).  
The multichannel generalization\cite{Ross:1961jlg,deSwart:1962,Burke:2011}
of the effective range expansion is
\beq
 K^{{-1}(J)}_{L'S'a';\ LSa}(\Ecm)=q_{{\rm cm},a'}^{-L'-\frac{1}{2}}
 \ {\widetilde{K}}^{{-1}(J)}_{L'S'a';\ LSa}(\Ecm)
  \ q_{{\rm cm},a}^{-L-\frac{1}{2}},
\label{eq:Keffrange}
\eeq
where ${\widetilde{K}}^{{-1}(J)}_{L'S'a';\ LSa}(\Ecm)$ is a real, symmetric, and often
analytic function of the center-of-momentum energy $\Ecm$. 
For a given total momentum $\Pvec=(2\pi/L)\dvec$ in a spatial $L^3$ volume
with periodic boundary conditions, where $\dvec$ is a vector of integers, 
we determine the total lab-frame energy $E$ for a
two-particle interacting state in our lattice QCD simulations.  If the 
masses of the two particles in decay channel $a$ are $m_{1a}$ and $m_{2a}$, we boost to the 
center-of-mass frame and define
\begin{eqnarray}
   \Ecm &=& \sqrt{E^2-\Pvec^2},\qquad
   \gamma = \frac{E}{\Ecm},\qquad
   \qqa = \frac{1}{4} \Ecm^2
   - \frac{1}{2}(m_{1a}^2+m_{2a}^2) + \frac{(m_{1a}^2-m_{2a}^2)^2}{4\Ecm^2},\nonumber\\
   u_a^2&=& \frac{L^2\qqa}{(2\pi)^2},\qquad
 \svec_a = \left(1+\frac{(m_{1a}^2-m_{2a}^2)}{\Ecm^2}\right)\dvec.
\end{eqnarray}
The total lab-frame energy $E$ is related to the scattering $K$-matrix
through the quantization condition:
\beq
\det(1-B^{(\Pvec)}\widetilde{K})=\det(1-\widetilde{K}B^{(\Pvec)})=0,\qquad
  \det(\widetilde{K}^{-1}-B^{(\Pvec)})=0.
\label{eq:quant}
\eeq
The \textit{box matrix} is given by
\beqs
 && \me{J'm_{J'}L'S'a'}{B^{(\Pvec)}}{Jm_JLS a} =
-i\delta_{a'a}\delta_{S'S} \ q_{{\rm cm},a}^{L'+L+1}\ W_{L'm_{L'};\ Lm_L}^{(\Pvec a)}  \nn\\
&&\qquad\qquad \times\langle J'm_{J'}\vert L'm_{L'},Sm_{S}\rangle
\langle Lm_L,Sm_S\vert Jm_J\rangle.
\label{eq:Bmatdef}
\eeqs
This box matrix $B^{(\Pvec)}$ is Hermitian for $q_{{\rm cm},a}^2$ real,
and the determinants in Eq.~(\ref{eq:quant}) are real.  
The $\langle j_1m_1 j_2m_2\vert JM\rangle$ are the familiar Clebsch-Gordan coefficients,
and the $W^{(\Pvec a)}$ matrix elements are given by
\[
-iW^{(\Pvec a)}_{L'm_{L'};\ Lm_L} 
= \sum_{l=\vert L'-L\vert}^{L'+L}\sum_{m=-l}^l
   \frac{ {\cal Z}_{lm}(\svec_a,\gamma,u_a^2) }{\pi^{3/2}\gamma u_a^{l+1}}
\sqrt\frac{(2{L'}+1)(2l+1)}{(2L+1)}\langle {L'} 0,l 0\vert L 0\rangle
\langle {L'} m_{L'},  l m\vert  L m_L\rangle.
\]
The Rummukainen-Gottlieb-L\"uscher (RGL) shifted zeta functions are evaluated
using
\beqs
   {\cal Z}_{lm}(\svec,\gamma,u^2)&=&\sum_{\nvec\in \mathbb{Z}^3}
  \frac{{\cal Y}_{lm}(\zvec)}{(\zvec^2-u^2)}e^{-\Lambda(\zvec^2-u^2)}
 +\delta_{l0}\frac{\gamma\pi}{\sqrt{\Lambda}} F_0(\Lambda u^2)
\nn\\
 &+&\frac{i^l\gamma}{\Lambda^{l+1/2}} \int_0^1\!\!dt 
\left(\frac{\pi}{t}\right)^{l+3/2}\! e^{\Lambda t u^2}
\sum_{\nvec\in \mathbb{Z}^3\atop \nvec\neq 0}
e^{\pi i \nvec\cdot\svec}{\cal Y}_{lm}(\wvec)
\  e^{-\pi^2\wvec^2/(t\Lambda)},
\label{eq:zaccfinal}
\eeqs
where $\zvec= \nvec -\gamma^{-1} \bigl[\textstyle\frac{1}{2}
+(\gamma-1)s^{-2}\nvec\cdot\svec \bigl]\svec$ and
$\wvec=\nvec - (1  - \gamma) s^{-2}
 \svec\cdot\nvec\svec$, the spherical harmonic polynomials are given by
${\cal Y}_{lm}(\xvec)=\vert \xvec\vert^l\ Y_{lm}(\widehat{\xvec})$,
and 
\begin{equation}
  F_0(x) =  -1+\frac{1}{2}
\int_0^1\!\! dt\ \frac{e^{tx}-1 }{t^{ 3/2}}.
\end{equation}
We choose $\Lambda\approx 1$ which allows sufficient
convergence speed of the summations. 

To make practical use of the determinants in Eq.~(\ref{eq:quant}), we change to
a block-diagonal basis and truncate in orbital angular momentum.  Matrices corresponding
to symmetry operations in the little group of $\Pvec$ commute with the box matrix,
leading to block-diagonal basis states
\beq
  \ket{\Lambda\lambda n JLS a}= \sum_{m_J} c^{J(-1)^L;\,\Lambda\lambda n}_{m_J} 
 \ket{Jm_JLS a},
\eeq
where $\Lambda$ denotes an irrep of the little group, $\lambda$ labels the row of the
irrep, and $n$ is an occurrence index.  The transformation coefficients depend on $J$ and 
$(-1)^L$, but not on $S,a$.  In this block-diagonal basis, the box matrix and the
$\widetilde{K}$ matrix for $(-1)^{L+L'}=1$ have the forms
\beqs
 \me{\Lambda'\lambda' n'J'L'S' a'}{B^{(\Pvec)}}{\Lambda\lambda nJLS a}
 &=& \delta_{\Lambda'\Lambda}\delta_{\lambda'\lambda}\delta_{S'S}
 \delta_{a'a}\ B^{(\Pvec\Lambda_B Sa)}_{J'L'n';\ JLn}(\Ecm),\\
 \me{\Lambda'\lambda' n'J'L'S' a'}{\widetilde{K}}{\Lambda\lambda nJLS a}
&=& \delta_{\Lambda'\Lambda}\delta_{\lambda'\lambda}\delta_{n'n} \delta_{J'J}
\ \widetilde{K}^{(J)}_{L'S'a';\ LS a}(\Ecm).
\eeqs

The quantization condition in Eq.~(\ref{eq:quant}) is a single relation
between an energy $E$ determined in finite-volume and the entire
$K$-matrix.  When multiple partial waves or multiple
channels are involved, this relation is clearly not sufficient
to extract all of the $K$-matrix elements at the single energy $E$.  The best way to 
proceed is to approximate the $K$-matrix elements using physically motivated 
functions of the energy $\Ecm$ involving a handful of parameters.  Values
of these parameters can then be estimated by appropriate fits
using a sufficiently large number of different energies.

\section{\boldmath The $\Delta$ Resonance}

\begin{figure}[t]
\begin{center}
 \includegraphics[width=0.45\textwidth]{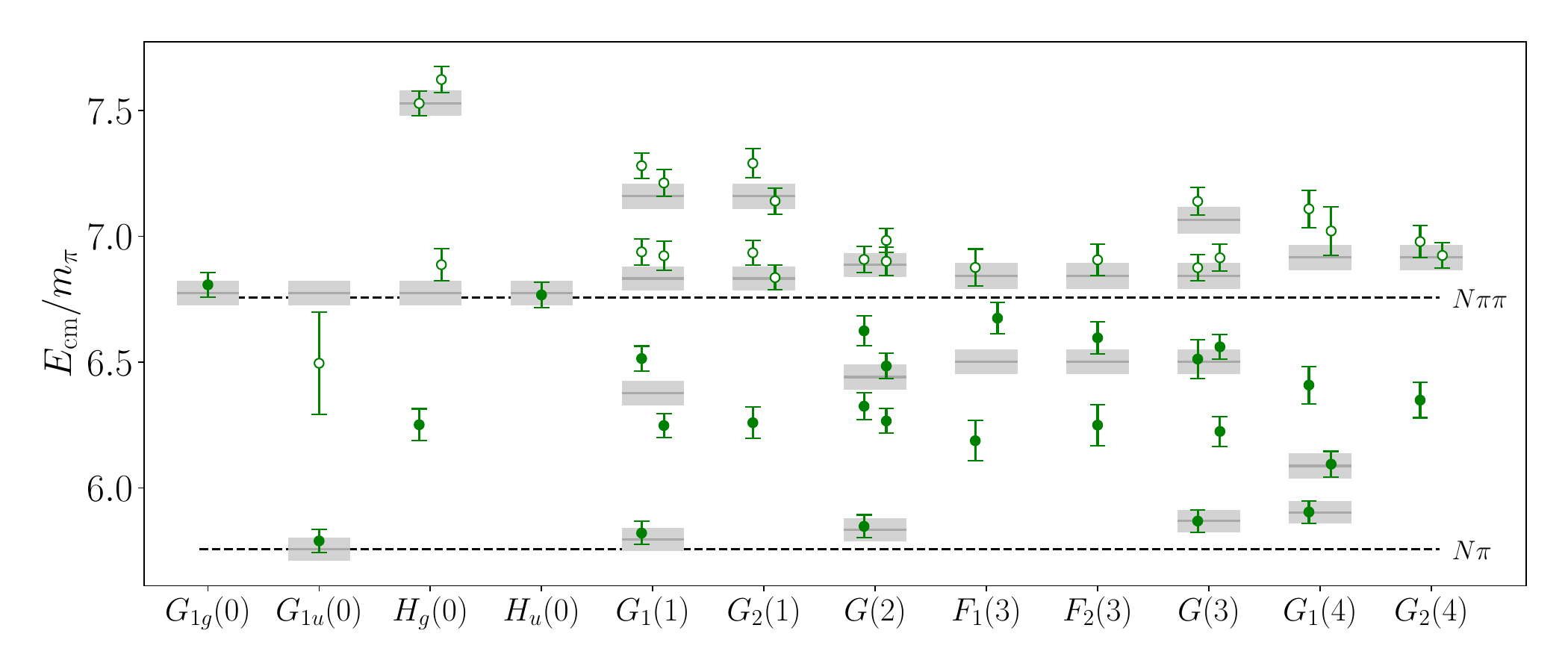}
 \includegraphics[width=0.45\textwidth]{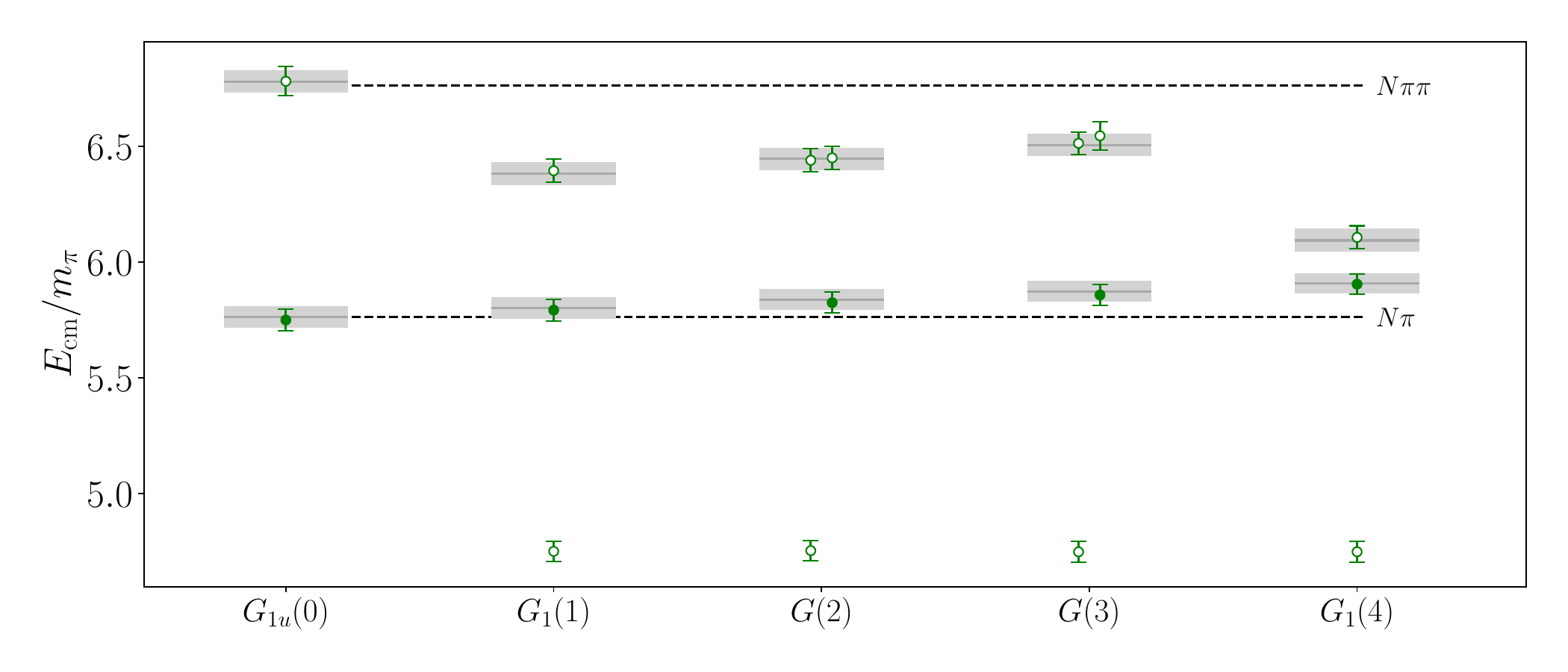}
\end{center}
\caption{The low-lying $I = 1/2$ and $I = 3/2$ nucleon-pion spectra in the 
center-of-momentum frame on the D200 ensemble as energies over the pion mass
from Ref.~\cite{Bulava:2022vpq}. Each column corresponds to a 
particular irrep $\Lambda$ of the little group of total momentum 
$\Pvec^2 = (2\pi/L)^2 \bm{d}^2$, denoted $\Lambda(\bm{d}^2)$. Dashed lines 
indicate the boundaries of the elastic region. Solid lines and shaded regions
indicate non-interacting $N \pi$ levels and their associated
statistical errors.
\label{fig:deltaspectrum}}
\end{figure}

\begin{figure}[b]
\begin{center}
\includegraphics[width=0.45\linewidth]{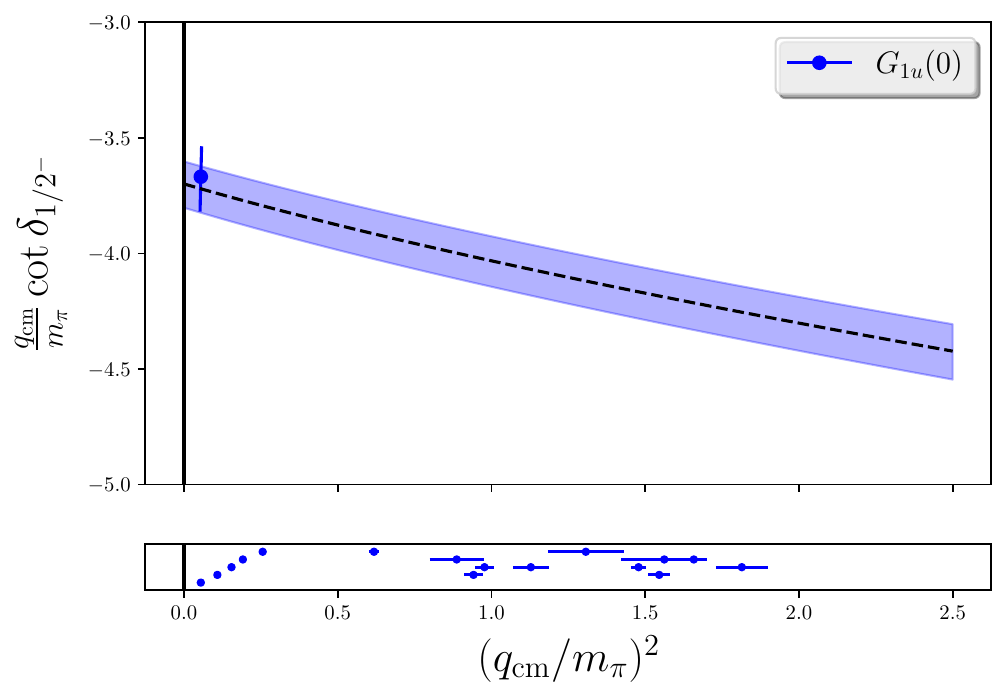}
\includegraphics[width=0.45\linewidth]{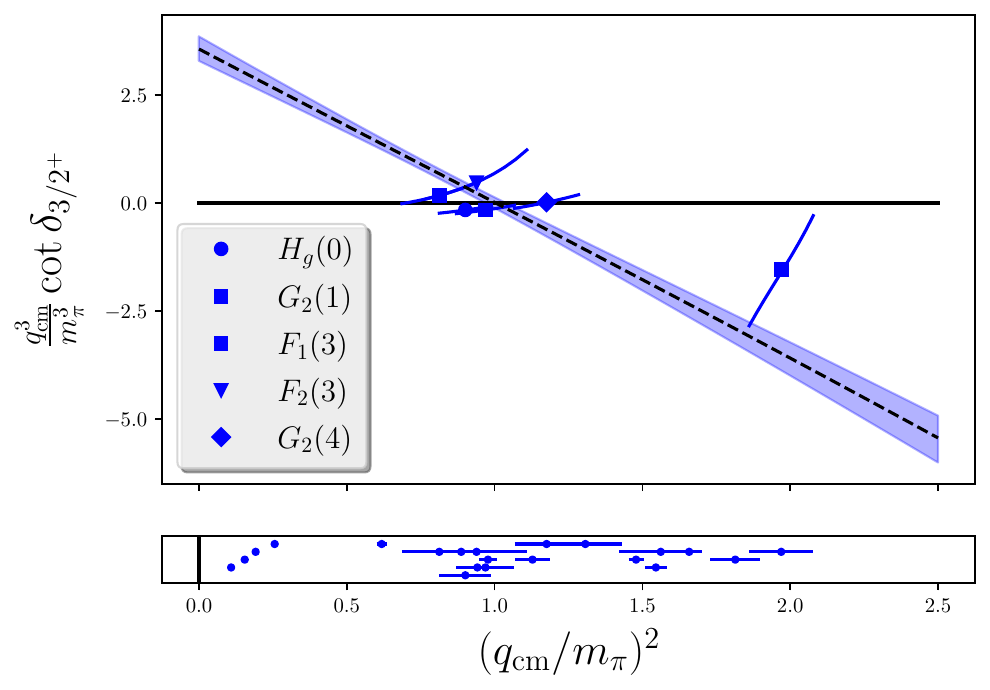}
\includegraphics[width=0.45\linewidth]{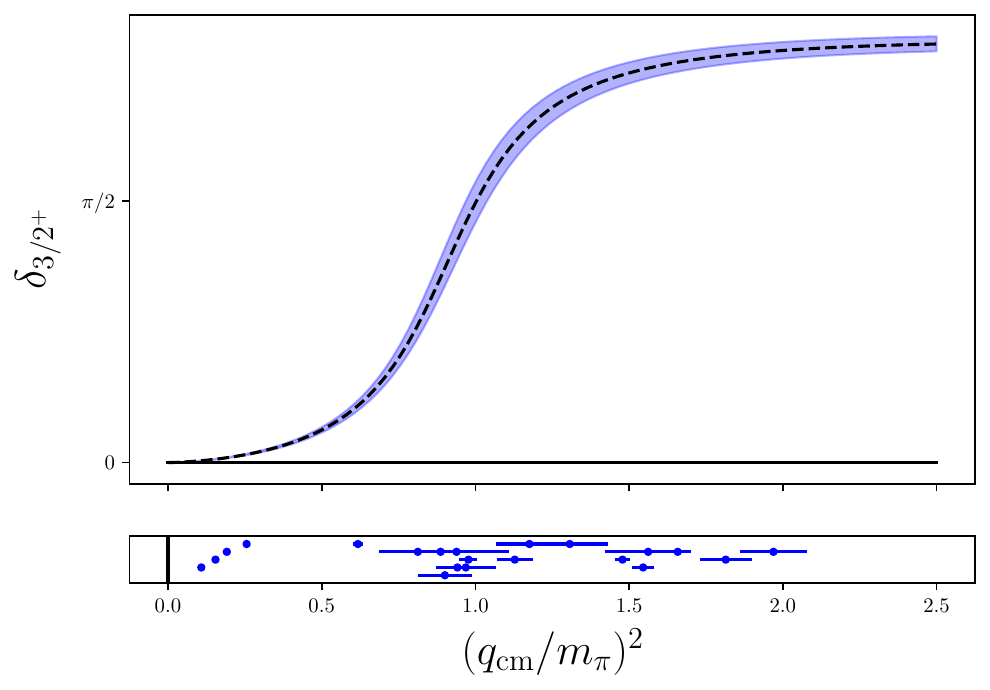}\quad
\raisebox{2mm}{\includegraphics[width=0.45\linewidth]{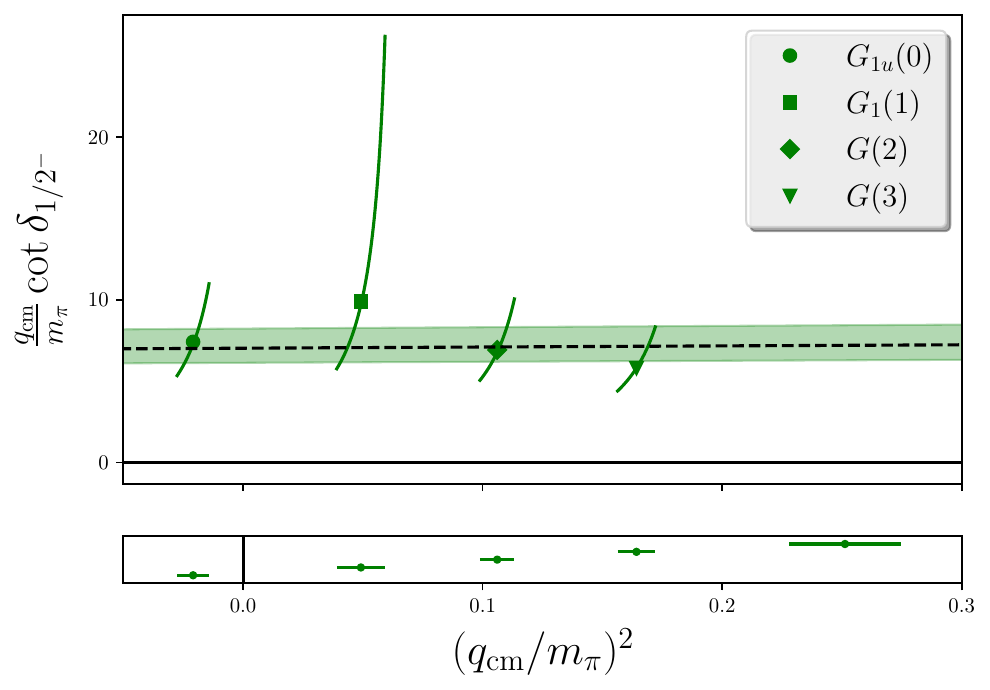}}
\end{center}
\vspace*{-8mm}
\caption{$N\pi$ scattering phase shifts from Ref.~\cite{Bulava:2022vpq}
for $I=\frac{3}{2}$: $s$-wave (top left), $p$-wave (top right)
in their cotangent form multiplied with threshold momentum factors.  The $p$-wave phase shift itself
 is shown in the bottom left.
 Similarly, $N\pi$ scattering phase shifts for $I=\frac{1}{2}$: $s$-wave (bottom right).
 Lower panels indicate all of the energies used in the fits to obtain the phase shifts in the
 top panels. \label{fig:delta2}}
\end{figure}

The $\Delta$ resonance is an important feature of nucleon-pion scattering.
In Ref.~\cite{Bulava:2022vpq}, our most recent study of $N\pi$ scattering at 
$m_\pi\sim 200~{\rm MeV}$  was presented.  Correlators related to meson-baryon 
and baryon-baryon scattering were computed using 2000 configurations with four 
source times of the CLS D200 ensemble, which employs a $64^3\times 128$ lattice 
with spacing $a\sim 0.065~\rm{fm}$ and open boundary conditions in time.  The 
quark masses are tuned such that $m_\pi\sim 200~\rm{MeV}$ and 
$m_K\sim 480~{\rm MeV}$. Results for finite-volume energies
obtained are shown in Fig.~\ref{fig:deltaspectrum}.  

For the 
$(2J, L) = (3, 1)$ wave, energies in the $H_{g}(0)$, $G_2(1)$, $F_1(3)$, $G_2(4)$
irrep were used.  In each irrep label, the integer in parentheses indicates
$\bm{d}^2$, for total momentum $\Pvec=2\pi\bm{d}/L$. The $G_{1u}(0)$ irrep
gives the $(1, 0)$ wave, the irreps used with $s$- and $p$-wave mixing
were $G_1(1)$, $G(2)$, $G_1(4)$. The scattering phase shifts
obtained from the finite-volume energies using the L\"{u}scher quantization
condition are shown in Fig.~\ref{fig:delta2}.  For the $\Delta$ mass and
Breit-Wigner width parameter $g_{\Delta,\rm{BW}}$, as well as the
scattering lengths, the following results were obtained:
\begin{equation}
 m_\Delta/m_\pi = 6.290(18),\quad g_{\Delta,\rm{BW}}=14.7(7),\quad
 m_{\pi}a_0^{3/2} = -0.2735(81),\quad  m_{\pi}a_0^{1/2} = 0.142(22).
\end{equation}
The amplitudes are well-described by the effective range expansion, and
a comparison to chiral perturbation theory was made.

\begin{figure}
\begin{center}
\includegraphics[width=4.0in]{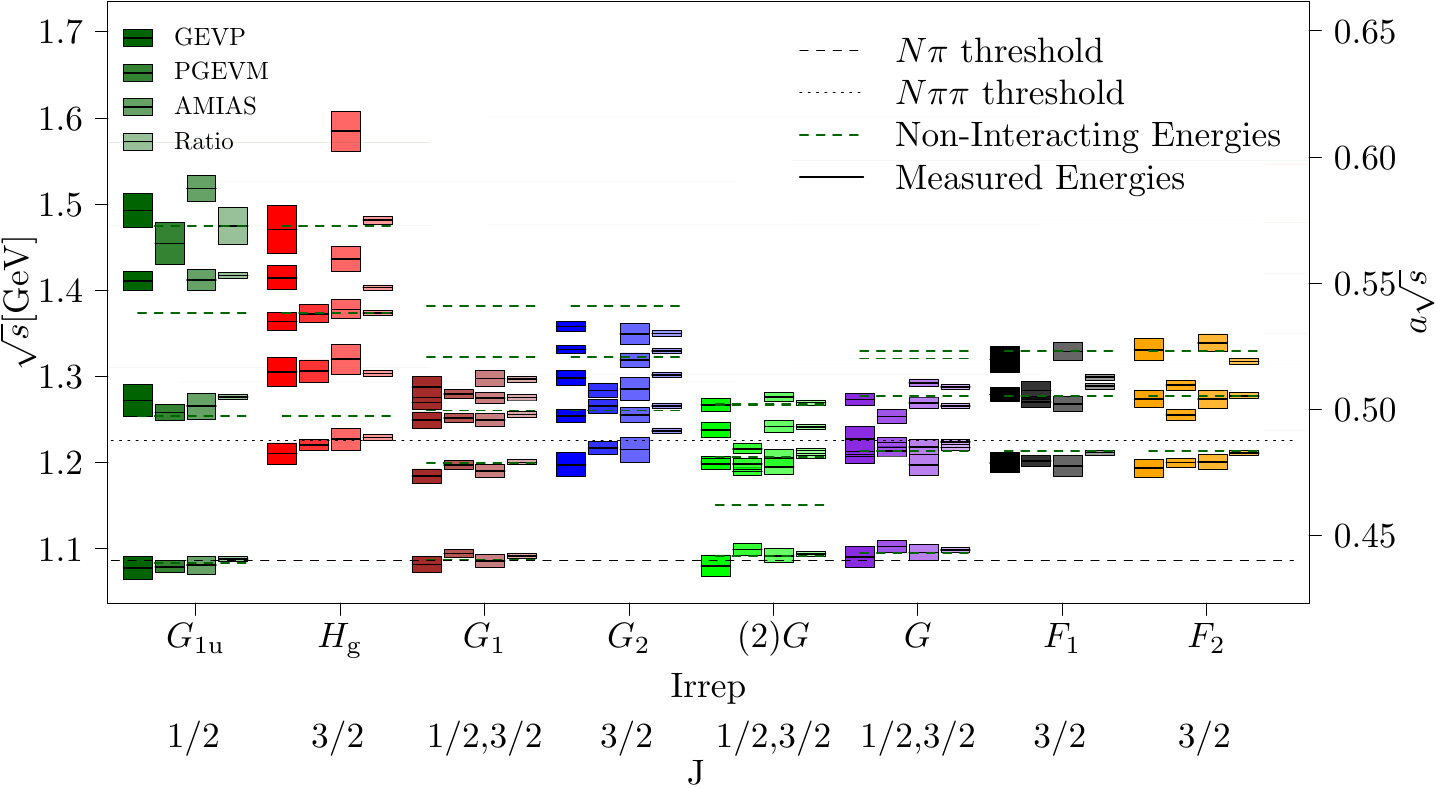}\\[4mm]
\includegraphics[width=3.0in]{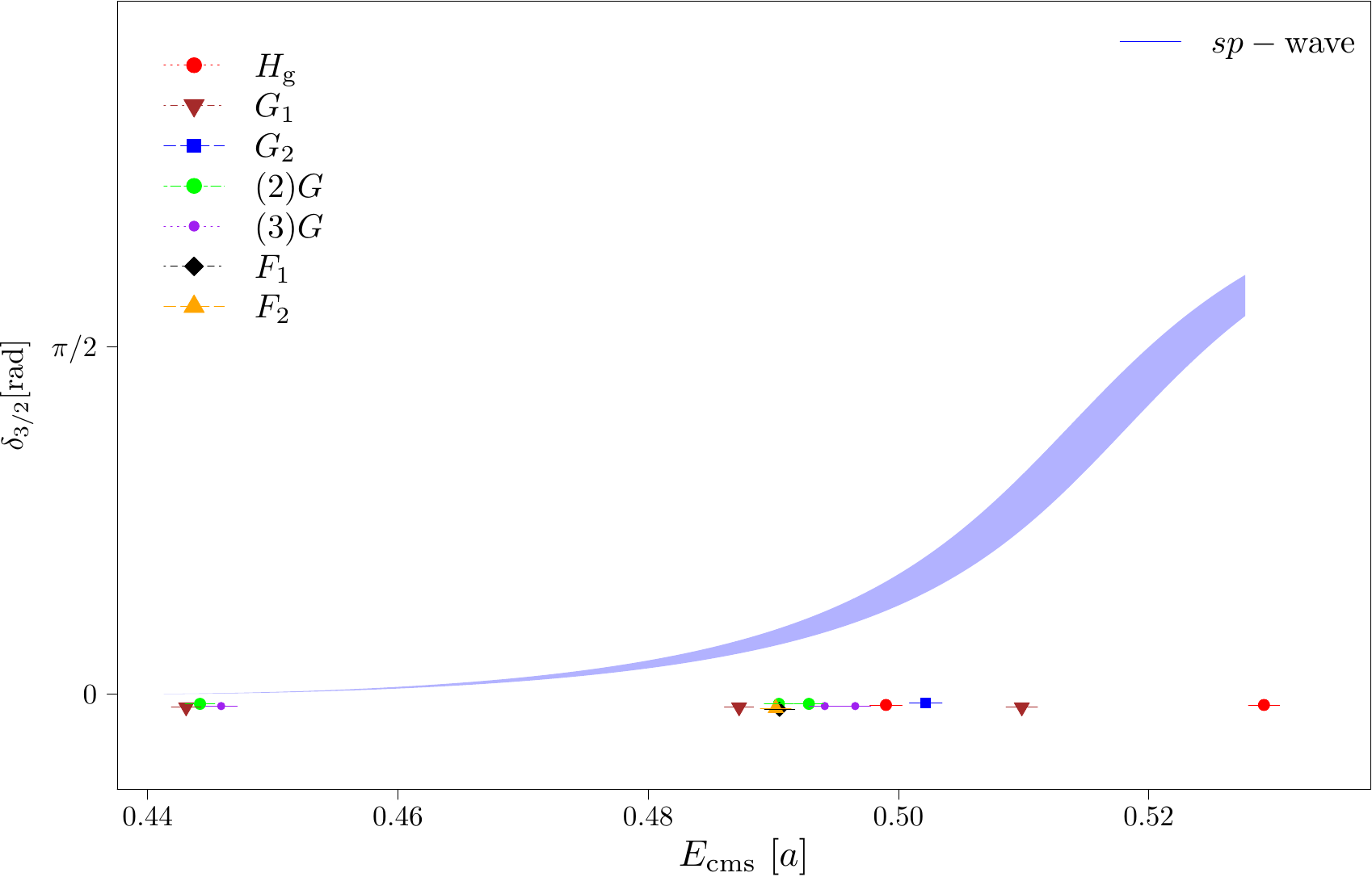}
\end{center}
\caption{(Top) The $\pi N$ interacting two-hadron energy levels obtained in 
Ref.~\cite{Alexandrou:2023elk}.  Box heights indicate estimated uncertainties.
Horizontal dashed/dotted lines show various thresholds, as indicated by the legend.
Noninteracting energies are shown by the green, thicker dashed lines.
(Bottom) The $P$-wave scattering phase-shift as a function of the invariant
mass $E_{\rm cm} = \sqrt{s}$. The error band is determined using jackknife
resampling. The points with horizontal error bars show each fitted
energy level included its jackknife error bar.
\label{fig:physDelta}}
\end{figure}

A study of the $\Delta$ resonance at the physical point (with quark masses 
set to give the physical pion and kaon masses) and lattice spacing $a= 0.08~\rm{fm}$  
was recently presented in Ref.~\cite{Alexandrou:2023elk}.  Their finite-volume
spectrum and scattering phase shift are shown in Fig.~\ref{fig:physDelta}.
Low three-particle thresholds were a problem in
this study.  The $\Delta$ resonance mass and width were found to be
\begin{eqnarray*}
 M_R &=& 1269\,(39)_{\rm{Stat.}}(45)_{\rm{Total}}~{\rm MeV},\\
\Gamma_R &=& 144\,(169)_{\rm{Stat.}}(181)_{\rm{Total}}~{\rm MeV}.
\end{eqnarray*}

\section{\boldmath Two-Pole Nature of Scattering near the $\Lambda(1405)$}

\begin{figure}[t]
\begin{center}
\includegraphics[width=0.53\columnwidth]{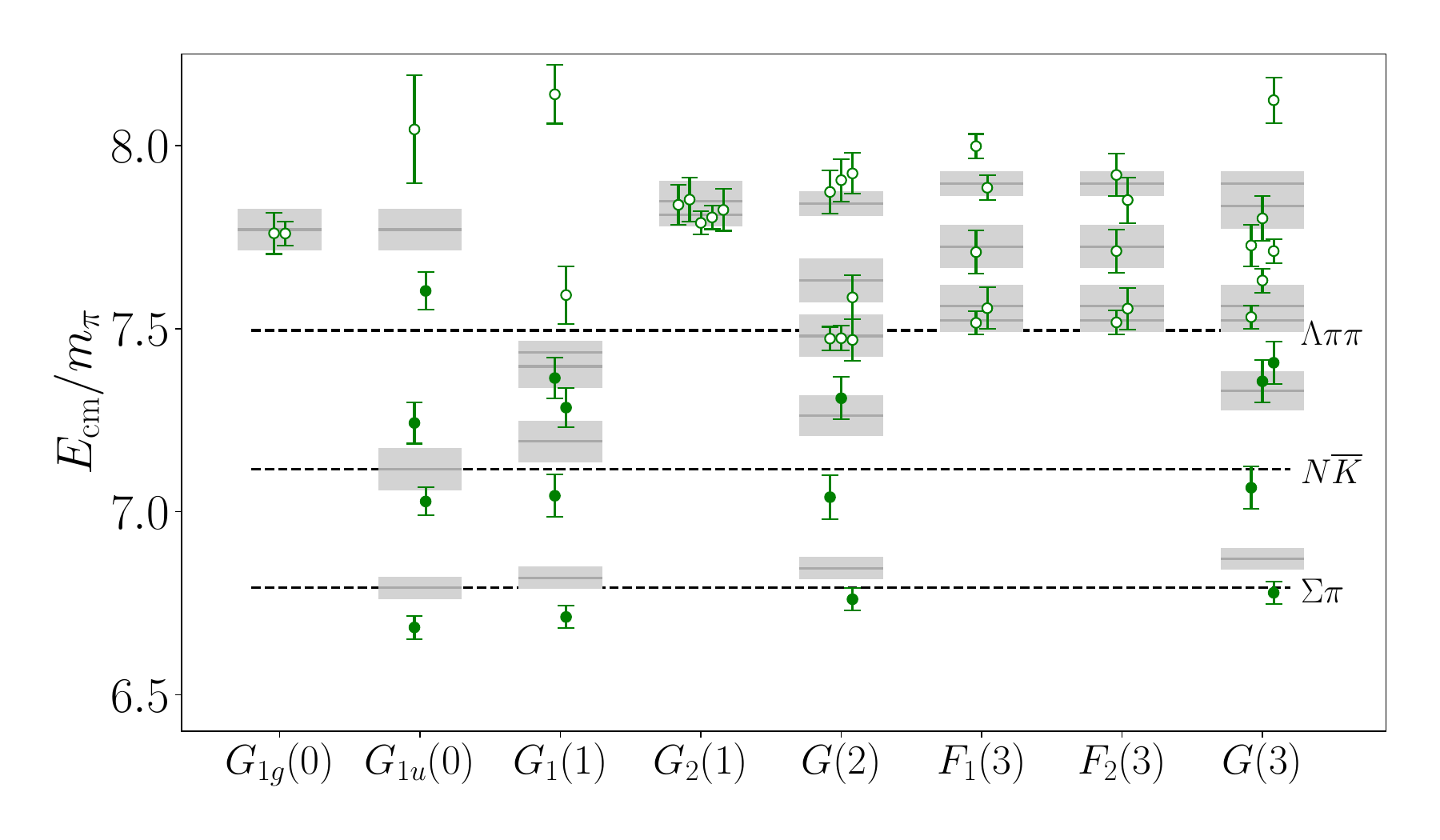}
\includegraphics[width=0.43\columnwidth]{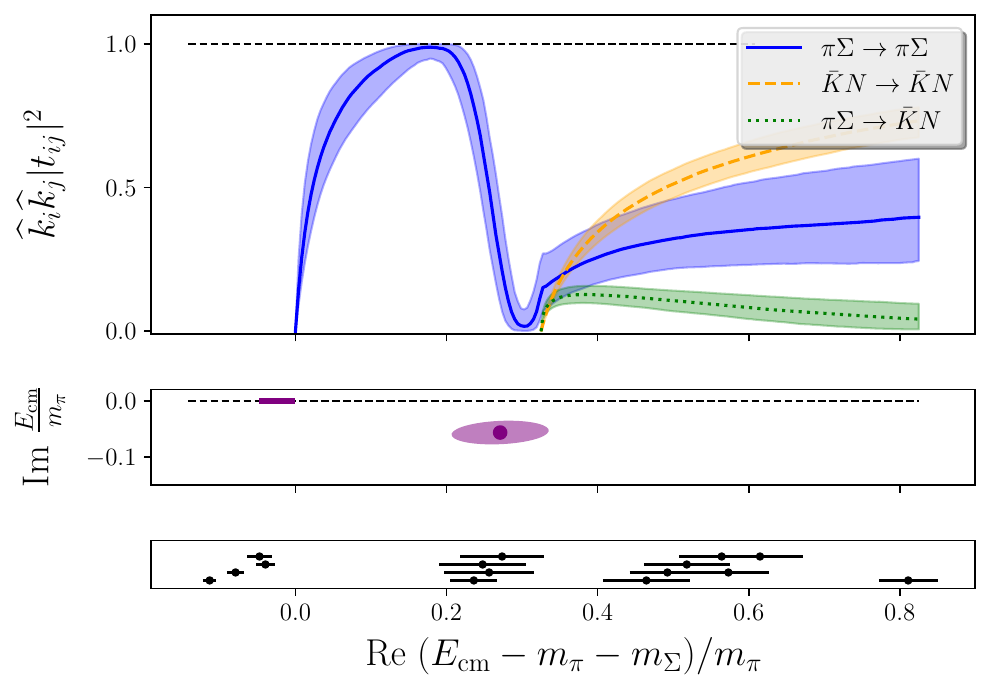}
\includegraphics[width=0.48\columnwidth]{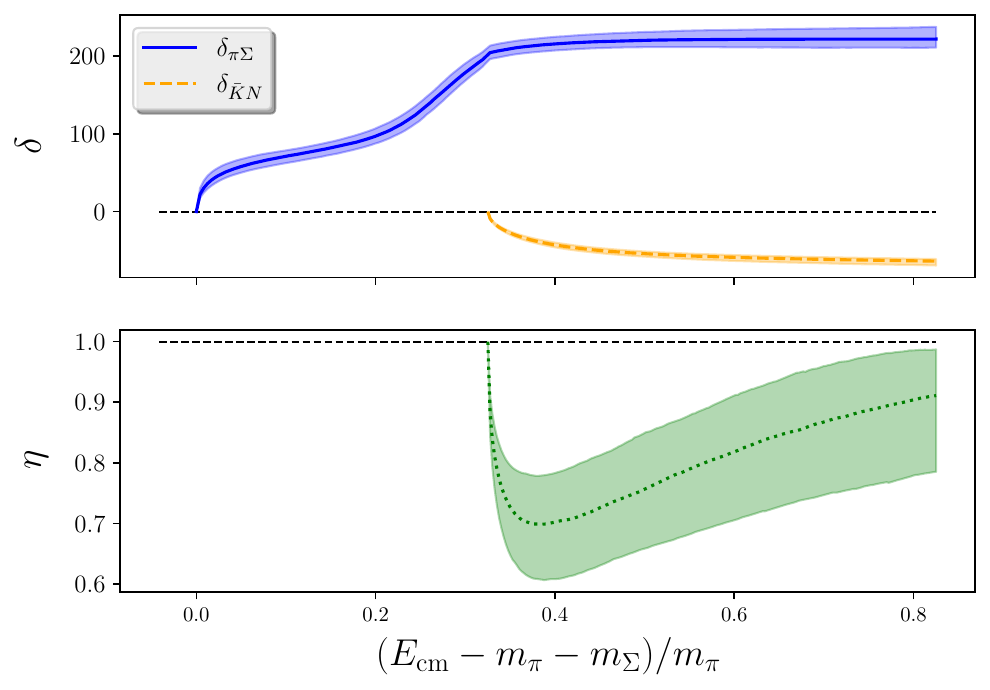}\hspace*{15mm}
\includegraphics[width=0.35\textwidth]{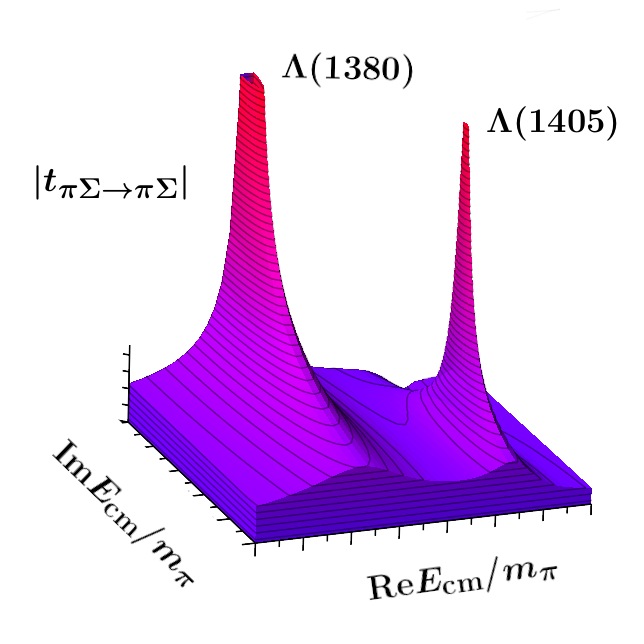}
\end{center}
\vspace*{-5mm}
\caption{(Top left) Finite volume energy spectrum involving interacting $\Sigma\pi$ and 
$N \overline{K}$ states as ratios over the pion mass
from Refs.~\cite{BaryonScatteringBaSc:2023zvt,BaryonScatteringBaSc:2023ori}.  
Green symbols are our results,
gray bands show non-interacting energies.  Labels on the horizontal axis show the irreps 
$\Lambda(\bm{d}^2)$ for lab-frame total momenta $\bm{P}=(2\pi/L)\bm{d}$, where $\bm{d}$ 
is a three-vector of integers and the lattice spatial volume is $L^3$.
(Top right) Upper panel shows the isoscalar, strangeness $-1$, $i\rightarrow j$
transition amplitudes squared for $i,j=\Sigma\pi,N\overline{K}$;
 middle panel shows positions of the $S$-matrix poles
in the complex center-of-mass energy plane on the sheet closest to the physical one;
bottom panel shows the finite-volume energies used in the fit.
(Bottom left) Inelasticity $\eta$ and phase shifts $\delta_{\pi\Sigma}$ and 
$\delta_{\overline{K}N}$.  (Bottom right) Three-dimensional plot of the 
$\Sigma\pi\rightarrow\Sigma\pi$ transition amplitude magnitude showing the two poles.
\label{fig:lambda}}
\end{figure}

In Refs.~\cite{BaryonScatteringBaSc:2023zvt,BaryonScatteringBaSc:2023ori}, 
our study of $\Sigma\pi$ and $N\bar{K}$ 
scattering in the $\Lambda(1405)$ energy region was presented.  Our results,
shown in Fig.~\ref{fig:lambda}, were obtained using the CLS D200 ensemble
with $m_\pi\sim 200~{\rm MeV}$.
This was the first lattice QCD study of this system to include both single-hadron and
all needed two-hadron operators to carry out a full coupled-channel analysis.
Our fits to the transition amplitudes revealed a two-pole structure, with
locations
\[
  E_1=1395(9)(2)(16)~{\rm MeV},\quad
  E_2=[1456(14)(2)(16)-i\, 11.7(4.3)(4)(0.1)]~{\rm MeV},
\]
with the first uncertainty being statistical, the second coming from our different
parametrizations of the amplitudes, and the third arising from scale setting.
A virtual bound state below the $\Sigma\pi$ threshold was found, as well as a
resonance pole below the $N\overline{K}$ threshold.  An effective range
expansion (ERE) with $\ell_{\rm max} = 0$ of the form
\beq
  \frac{E_{\rm cm}}{M_\pi} \tilde{K}_{ij} = A_{ij} + B_{ij} \Delta_{\pi \Sigma},
\qquad \Delta_{\pi \Sigma} = (E_{\rm cm}^2 - (M_\pi+M_\Sigma)^2)/(M_\pi+M_\Sigma)^2,
\eeq
where $A_{ij}$ and $B_{ij}$ are symmetric and real coefficients with $i$ and $j$ 
denoting either of the two scattering channels, provided the best description of 
the data, but several other parametrizations were
also used, including an ERE for $\tilde{K}^{-1}$, the form above with
the outer factor of $E_{\rm cm}$ removed, and a Blatt-Biedenharn form.
All forms with one pole were strongly disfavored.  The two-pole structure in
the $\Lambda(1405)$ region was first suggested in Ref.~\cite{Oller:2000fj}.

\section{Nucleon-Nucleon Scattering}

Given the signal-to-noise problem of baryons in lattice QCD, nucleon-nucleon studies
are particularly challenging.  A very early study of hadron scattering lengths
in the so-called quenched approximation was presented in Ref.~\cite{Fukugita:1994ve}.
Heroic attempts to study $NN$ systems in lattice QCD fully taking dynamical quarks
into account were carried out in the mid-2010's.  Studying $NN$ systems at the
$SU(3)$ flavor symmetric point for unphysically heavy quark masses was used as
a starting point to explore $NN$ scattering in lattice QCD.  Shallow bound states
in both $I=0$ and $I=1$ $NN$ systems were found in
Refs.~\cite{NPLQCD:2012mex,NPLQCD:2013bqy,Berkowitz:2015eaa}, among others.
Since the current computational techniques and current computing capabilities 
were not available then, the use of a single off-diagonal correlator was used
at that time to simplify these difficult calculations. Meanwhile, another study of
such systems using an alternative approach espoused by the HALQCD collaboration 
found that there were no bound states in either channel\cite{Inoue:2011ai}. This 
discrepancy in the early results was certainly an inauspicious beginning for 
$NN$ scattering in lattice QCD.

$NN$ scattering at the $SU(3)$ flavor symmetric point was later revisited in 
Ref.~\cite{Horz:2020zvv} using more up-to-date techniques, and no bound states 
were found in either the $I=0$ or $I=1$ $NN$ systems.  Furthermore, all other recent
studies\cite{Francis:2018qch,Amarasinghe:2021lqa,NNsu3,Green:2021qol,Green:2022rjj,Geng:2024dpk} 
using a Hermitian correlation matrix method have largely reached this
same conclusion.  This leads one to ask about the cause of the initial discrepancy.  
One suggestion that the use of a local 
hexaquark operator is needed in order to reliably extract the ground state energy in 
such systems, but two subsequent studies\cite{Amarasinghe:2021lqa,NNsu3} have 
convincingly shown that this is not the case, as will be described below.
Systematic effects, such as discretization errors,
cannot be ruled out, but these small effects are unlikely to be the cause of such a 
large discrepancy.  After a few years of investigating this discrepancy, a preponderous
of evidence suggests that the use of the off-diagonal correlator and plateau 
misidentification was the most likely culprit for the discrepancy.

\begin{figure}
\begin{center}
\raisebox{6mm}{\includegraphics[width=2.1in]{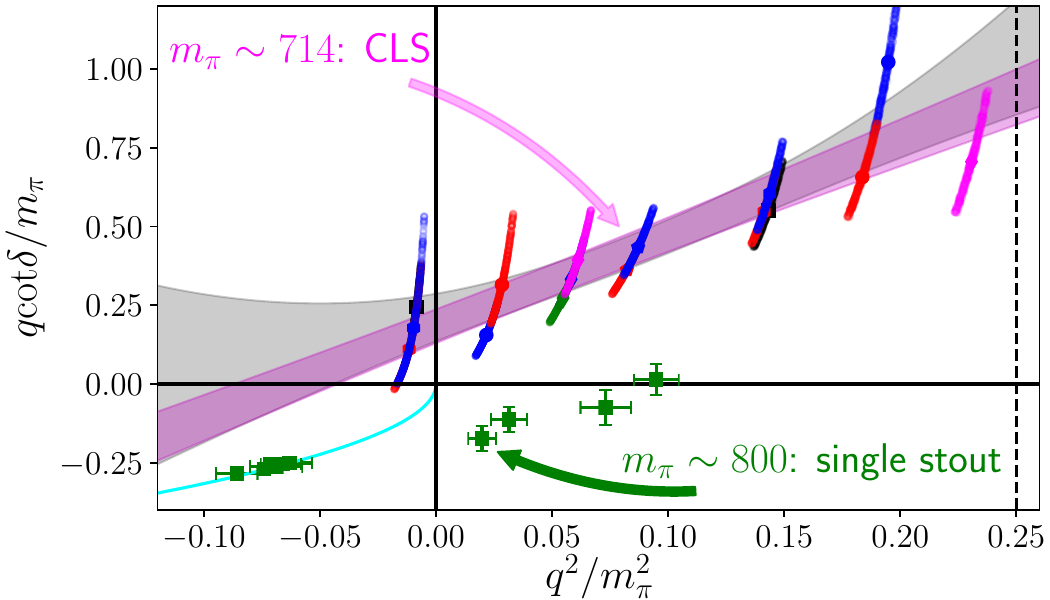}}
\includegraphics[width=1.8in]{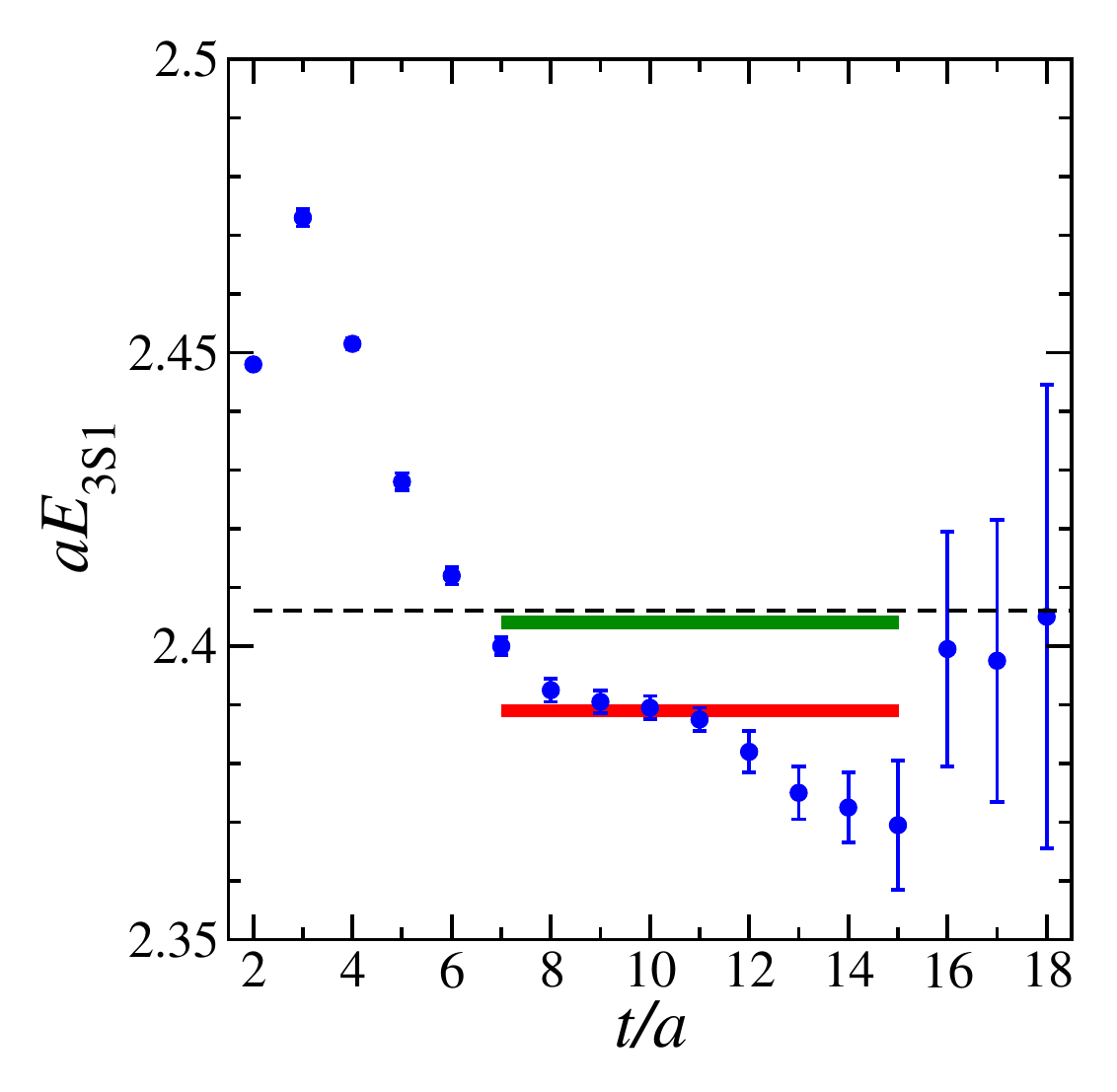}
\includegraphics[width=1.8in]{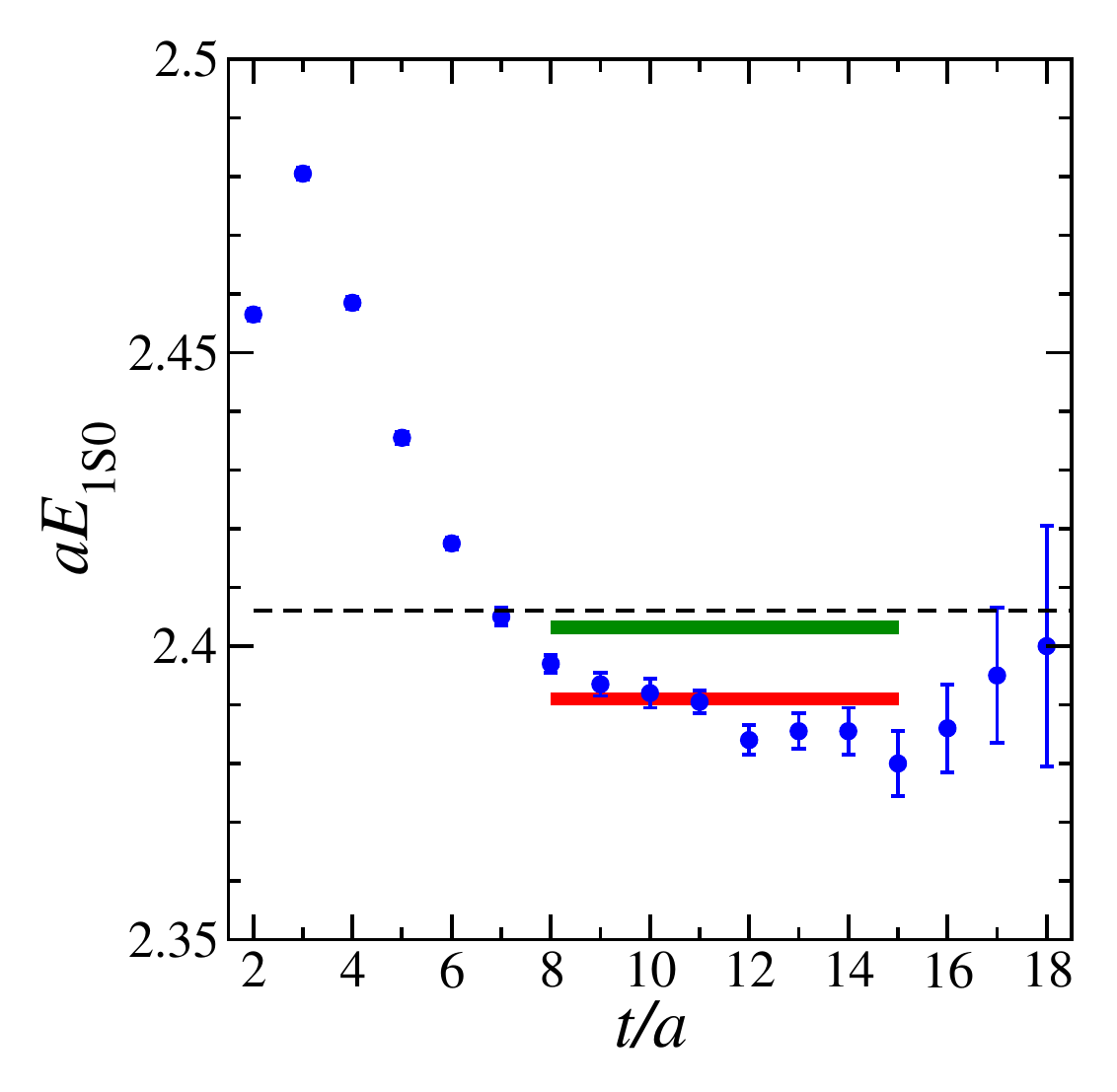}
\end{center}
\caption{(Left) Comparison of scattering phase shifts in the deuteron channel 
determined in Ref.~\cite{Horz:2020zvv}, shown as red, blue, magenta points and magenta
and gray bands, with phase shifts determined in Refs.~\cite{NPLQCD:2012mex,NPLQCD:2013bqy},
shown as green points.  (Center) Effective mass, shown as blue points, from 
Fig.~2 of Ref.~\cite{Beane:2017edf}
for the deuteron on a $48^3$ spatial lattice from an off-diagonal correlator involving a
local hexaquark operator at the source and a nucleon-nucleon operator of zero momentum 
at the sink. The horizontal red line indicates the energy extraction from Fig.~4 of
Ref.~\cite{NPLQCD:2012mex}, with the green line indicating 
approximately where the energy extraction from Ref.~\cite{Horz:2020zvv} would be for 
this $48^3$ lattice. (Right) Effective mass, similar to that in the center plot,
but for the dineutron.
\label{fig:NNdiscrepancy}}
\end{figure}

The crux of the discrepancy can be seen in Fig.~\ref{fig:NNdiscrepancy}. The
scattering phase shift $q\cot(\delta/m_\pi)$ for the deuteron is shown in the
left-hand plot of Fig.~\ref{fig:NNdiscrepancy}.  Early results from 
Refs.~\cite{NPLQCD:2012mex,NPLQCD:2013bqy}
which used an off-diagonal correlator are shown in green, suggesting a bound state, 
whereas more recent results from Ref.~\cite{Horz:2020zvv} which used a Hermitian correlation
matrix are shown by the red, blue, and magenta points with gray and magenta bands,
suggesting no bound state.  Note that Refs.~\cite{NPLQCD:2012mex,NPLQCD:2013bqy} used a 
tadpole-improved L\"uscher-Weisz gauge action and a stout-smeared clover fermion action
with lattice spacing 0.145~fm, while Ref.~\cite{Horz:2020zvv} used a tree-level improved
L\"uscher-Weisz gauge action and a non-perturbatively $O(a)$-improved clover Wilson fermion
action from CLS with lattice spacing 0.086~fm.  The quark masses
also differ, leading to a pion mass of 800~MeV in Refs.~\cite{NPLQCD:2012mex,NPLQCD:2013bqy} 
and about 710~MeV in Ref.~\cite{Horz:2020zvv}.  These lattice actions, spacings, and
volumes are similar enough that they are unlikely to be the cause of the large discrepancy
seen in the left-hand plot of Fig.~\ref{fig:NNdiscrepancy}.  

The discrepancy seems to boil down to a difference in energy extractions.  Effective masses
from Fig.~2 of Ref.~\cite{Beane:2017edf} for the deuteron (center plot) and dineutron
(right plot) are shown in Fig.~\ref{fig:NNdiscrepancy} for a $48^3$ spatial lattice.
A single off-diagonal correlator involving a local hexaquark operator at the source and
a nucleon-nucleon operator of zero momentum at the sink was used to determine these
effective masses, which are shown as blue circles with errors.   The horizontal red
lines indicate the energy extractions from Fig.~4 of Ref.~\cite{NPLQCD:2012mex}
with the vertical thickness indicating the statistical uncertainties.
The horizontal dashed lines indicate the energies of two non-interacting nucleons at rest
in the deuteron and dineutron cases.  The occurrences of the horizontal red lines well 
below the horizontal dashed lines, along with similar differences in other energy
extractions, essentially lead to the bound states observed, for example,
in the left plot.  The horizontal green boxes indicate approximately where the energy 
extractions from Ref.~\cite{Horz:2020zvv} would be for this $48^3$ lattice.  The lowest-lying
energies in Ref.~\cite{Horz:2020zvv} occur very slightly below the non-interacting energies.
Hence, the discrepancy essentially arises from the differences between the red and 
green lines, along with similar differences from other energy determinations.

\begin{figure}
\begin{center}
\begin{minipage}{3.8in}
\includegraphics[width=1.8in]{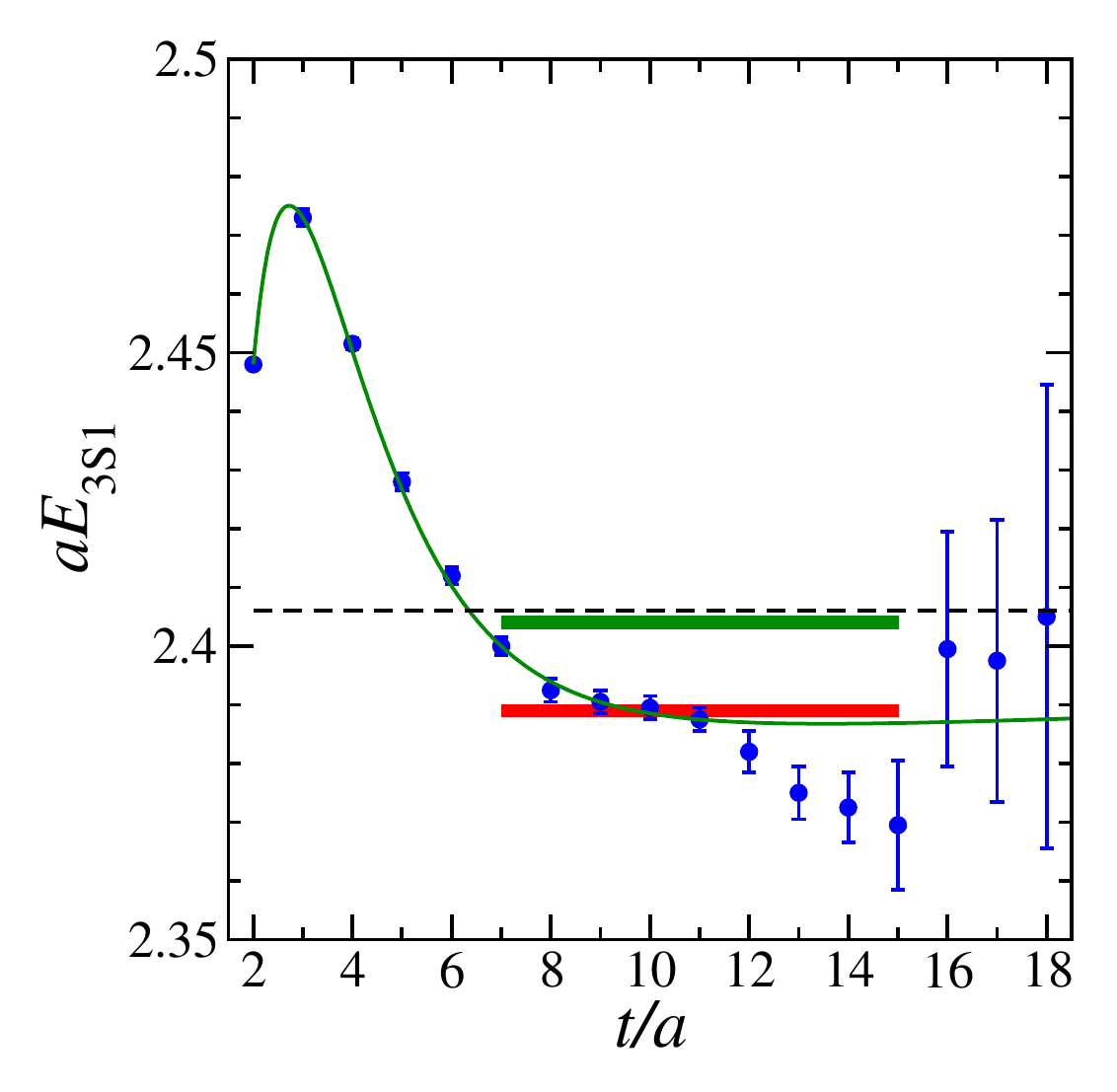}
\includegraphics[width=1.85in]{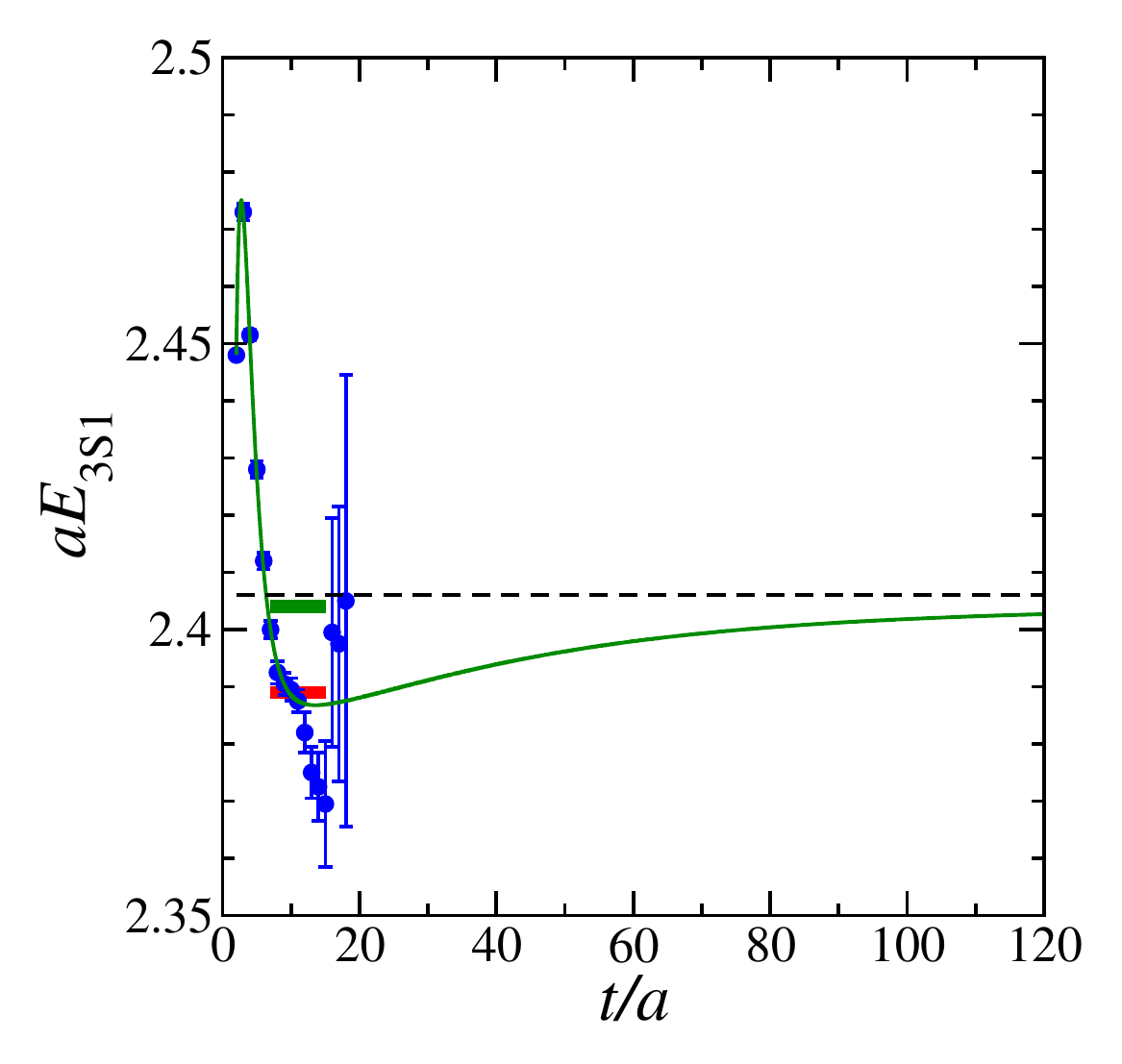}\\
\includegraphics[width=1.8in]{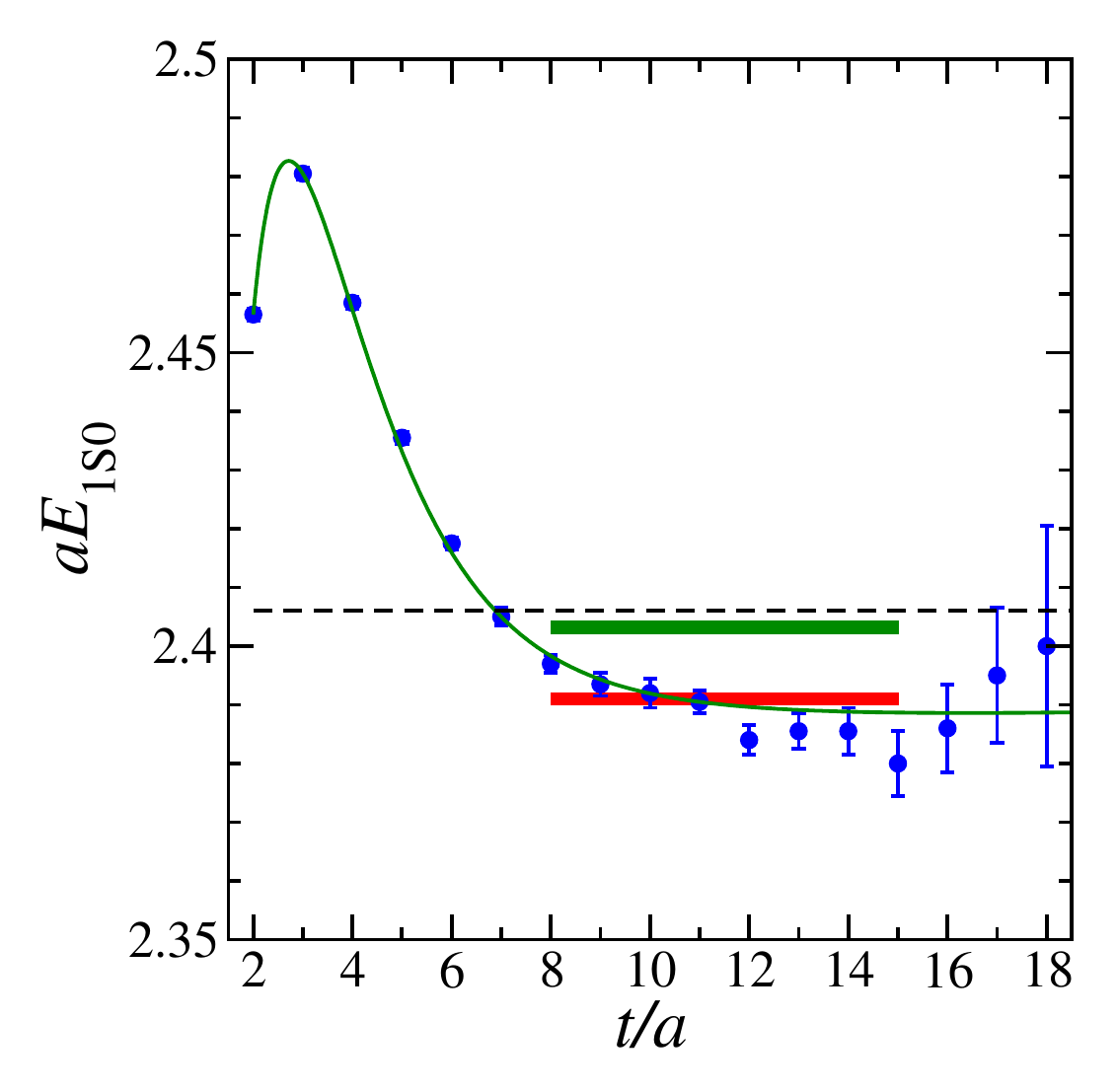}
\includegraphics[width=1.85in]{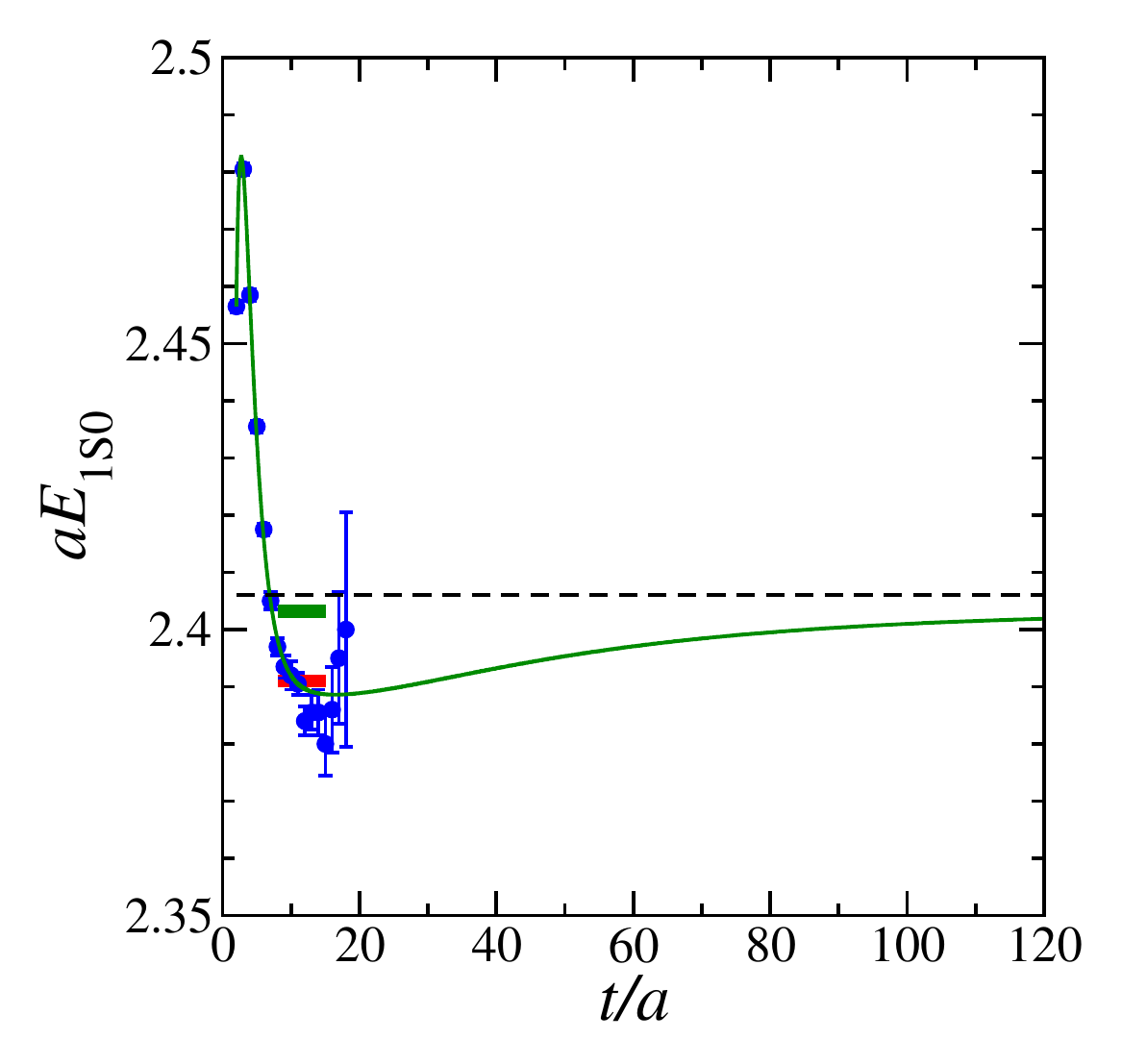}
\end{minipage}
\raisebox{-1.0in}{\includegraphics[width=2.0in]{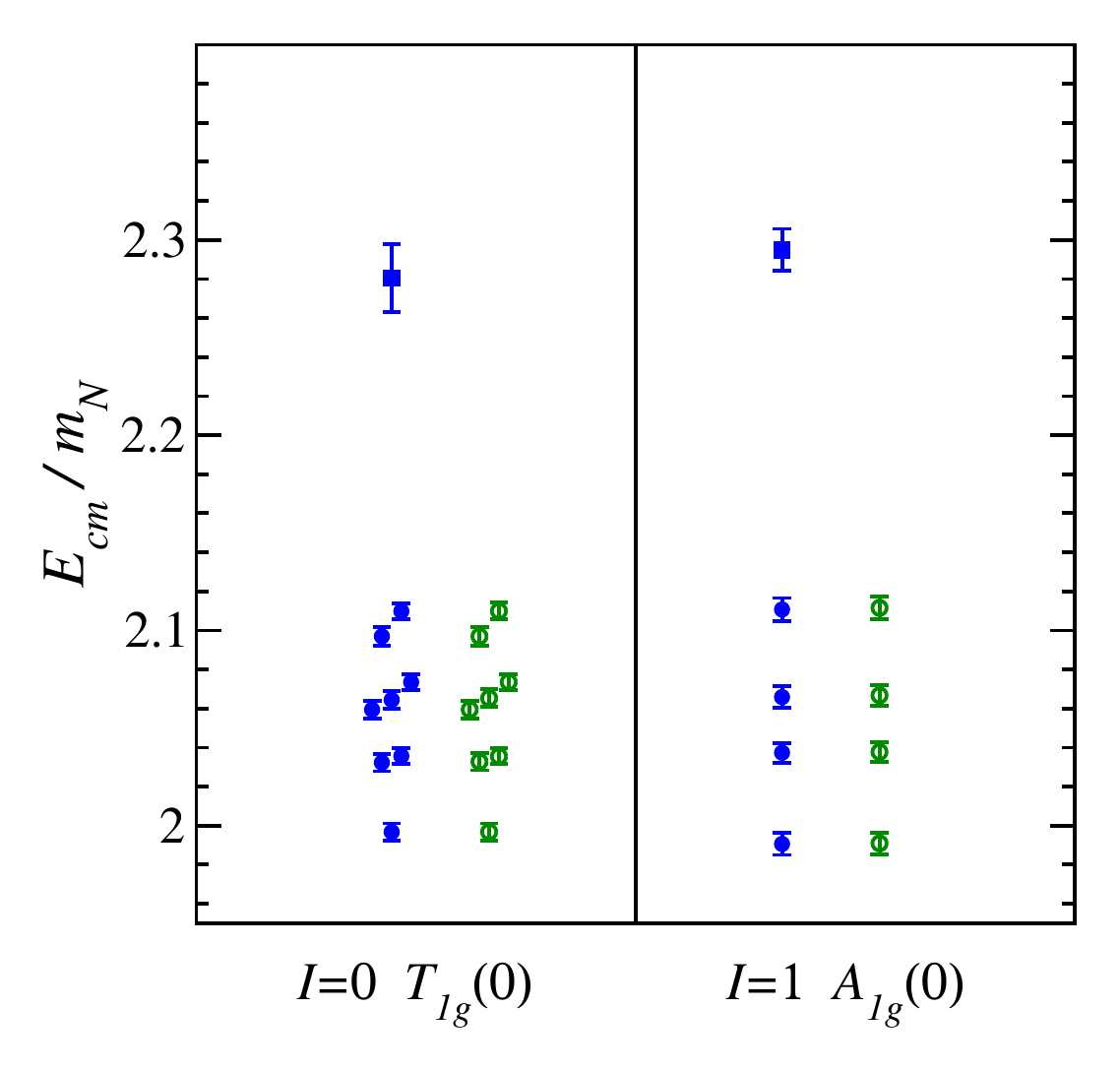}}
\end{center}
\caption{(Upper left) Same as the center plot of Fig.~\ref{fig:NNdiscrepancy},
but also includes the fit from Eqs.~(\ref{eq:NNfalsefit}), (\ref{eq:NNfalsegaps}), 
(\ref{eq:NNfalseAvalstriplet})
as a green curve. For temporal range $t/a=10\cdots 20$, the green curve shows a
remarkable but false plateau.  (Upper center) Same as the upper left plot in this figure, 
but continued to much larger temporal separations, showing the very slow approach to the
asymptotic limit.  (Lower left) Same as the right plot of Fig.~\ref{fig:NNdiscrepancy},
but also includes the fit from Eqs.~(\ref{eq:NNfalsefit}), (\ref{eq:NNfalsegaps}), 
(\ref{eq:NNfalseAvalssinglet})
as a green curve. (Lower center) Same as the lower left plot in this figure, but continued
to much larger temporal separations. (Right) Energy spectrum as ratios over nucleon 
mass for each isospin channel for total zero momentum. The blue circles and square indicate results
obtained using the entire correlation matrices, including the
hexaquark operator. Green circles show the energies obtained
using the correlation matrices excluding the hexaquark operators. 
The blue square in each channel indicates the energy
corresponding to a hexaquark-dominated level.
\label{fig:falseplateaux}}
\end{figure}

Temporal correlators in lattice QCD admit a spectral representation
of the form
\begin{equation}
     C_{ij}(t) = \sum_{n=0}^\infty Z_i^{(n)}Z_j^{(n)\ast} e^{-E_n t},
\end{equation}
ignoring negligible temporal wrap-around contributions.  For a diagonal 
$i=j$ correlator, the weights of the exponentials in the above spectral
representation are guaranteed to all be \textit{positive}
 \begin{equation}
     C_{ii}(t) = \sum_{n=0}^\infty \left\vert Z_i^{(n)}\right\vert^2 e^{-E_n t},
 \end{equation}
which greatly restricts the behavior of such correlators.  Off-diagonal 
$i\neq j$ correlators are not subject to such restrictions and can have 
\textit{negative} weights.  The initial rise of the blue points at small temporal
separation in Fig.~\ref{fig:NNdiscrepancy} is a major cause for concern, indicating
the unwelcome presence of large negative weights in the spectral representation. 
Furthermore, excited-state contamination in a
simple single off-diagonal correlator decays slowly as $e^{-(E_1-E_0)t}$,
where $E_0$ is the energy of the lowest-lying state and $E_1$ is the
energy of the second lowest-lying state.  Contamination in a diagonal
correlator obtained from a generalized eigenvalue problem optimization
decays much more quickly as $e^{-(E_N-E_0)t}$ for an $N\times N$ correlator 
matrix. Given the possibility of negative weights and the slow decay of 
excited-state contamination in a single off-diagonal correlator, the 
likelihood of plateau misidentification is uncomfortably high.

To illustrate how plateau misidentification can occur, consider a five-exponential
form for the off-diagonal correlator
 \begin{equation}
    C(t)=e^{-E_0 t}\Bigl(1+A_1e^{-\Delta_1 t}+A_2e^{-\Delta_2 t}
        +A_3e^{-\Delta_3 t}+A_4e^{-\Delta_4 t}\Bigr).
  \label{eq:NNfalsefit}
 \end{equation}
For the two lowest gaps, we take values that are expected for a $48^3$ spatial
lattice.  The other 2 gaps are set to be large enough to handle the observed
short-time behavior of the effective mass.  In particular, we take
\begin{equation}
   \Delta_1=0.025,\quad\Delta_2=\Delta_1+0.025,\quad\Delta_3=\Delta_2+0.5,\quad\Delta_4=\Delta_3+1.0,
   \label{eq:NNfalsegaps}
\end{equation}
then using the $E_0$ values shown by the green boxes, we can solve for the weights
$A_1,A_2,A_3,A_4$ using correlations at times  $t=2,3,7,11$.
For the deuteron $(I=0,\ ^3S_1)$, we find
\begin{equation}
    A_1 = -1.0483,\ A_2 = 0.4133,\  A_3 = 0.6495,\ A_4 = -1.7750,
\label{eq:NNfalseAvalstriplet}
\end{equation}
and for the dineutron $(I=1,\ ^1S_0)$, we obtain
\begin{equation}
     A_1 = -1.0986,\ A_2 = 0.4993,\  A_3 = 0.7127,\ A_4 = -1.9065.
\label{eq:NNfalseAvalssinglet}
\end{equation}
The resulting effective masses are shown as the green curves in the left and center
plots of Fig.~\ref{fig:falseplateaux}.  The green curves reproduce the observed behaviors
at small temporal separations and show amazingly flat plateaux-like behavior for
a temporal range from about $t/a=10$ to 20, as illustrated in the left-hand plots of
Fig.~\ref{fig:falseplateaux}.  However, the center plots display the behaviors
for larger temporal separations, showing how the approaches to the asymptotic limits
given by the green boxes are exceedingly slow.  These plots are for illustrative 
purposes only to show how this could happen; they do not prove that this did happen.
This illustration is similar to that presented in Ref.~\cite{Iritani:2018vfn}.

The right hand plot in Fig.~\ref{fig:falseplateaux} displays the role of the 
hexaquark operator in such calculations. For both the singlet and
triplet $NN$ channels, the blue points show the spectra obtained using all operators,
to be compared with the green points showing the spectra obtained using all operators
but excluding the hexaquark operators.  The inclusion of the hexaquark operators
has no effect on the low-lying energies extracted; an additional level is observed
far above all of the other levels.

\begin{figure}
\begin{center}
\includegraphics[width=2.8in]{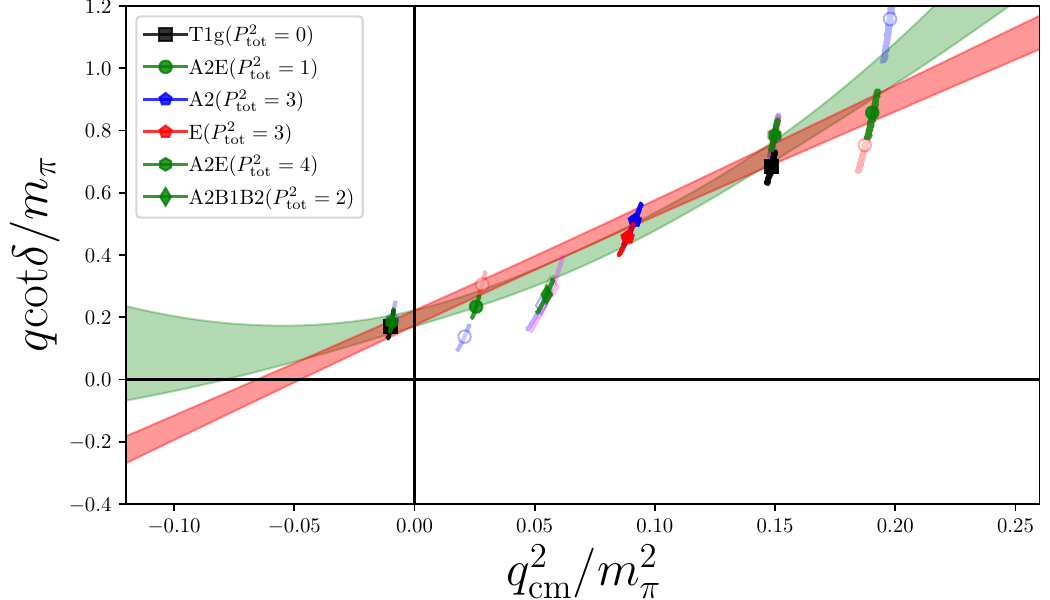}
\includegraphics[width=2.8in]{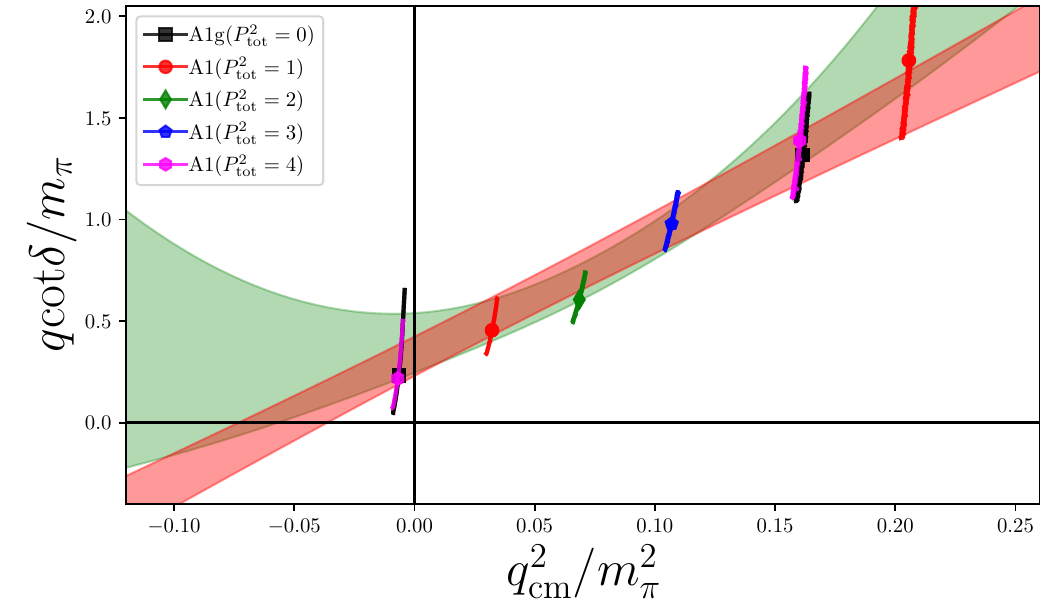}
\end{center}
\caption{Results for $(q/m_\pi)\ \cot(\delta)$ for $NN$ isosinglet $^3S_1$ and isotriplet 
$^1S_0$ scattering from Ref.~\cite{NNsu3} with improved statistics on the C103 ensemble 
from CLS at the $SU(3)$ flavor symmetric point with $m_\pi\sim 710$~MeV. 
The red bands are linear fits in $q^2$, and the green bands are quadratic fits.
\label{fig:NNsu3}}
\end{figure}

Our latest results\cite{NNsu3} for the $NN$ isosinglet $^3S_1$ and isotriplet 
$^1S_0$ scattering phases shifts with improved statistics on the C103 ensemble from 
CLS at the $SU(3)$ flavor symmetric point with $m_\pi\sim 710$~MeV are shown in 
Fig.~\ref{fig:NNsu3} and show no bound states.

\section{A Few Other Recent Studies}

\begin{figure}
\begin{center}
\includegraphics[width=2.8in]{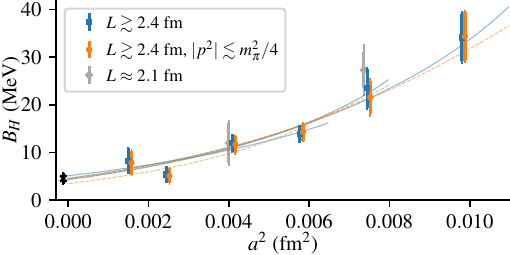}\hspace*{5mm}
\includegraphics[width=2.8in]{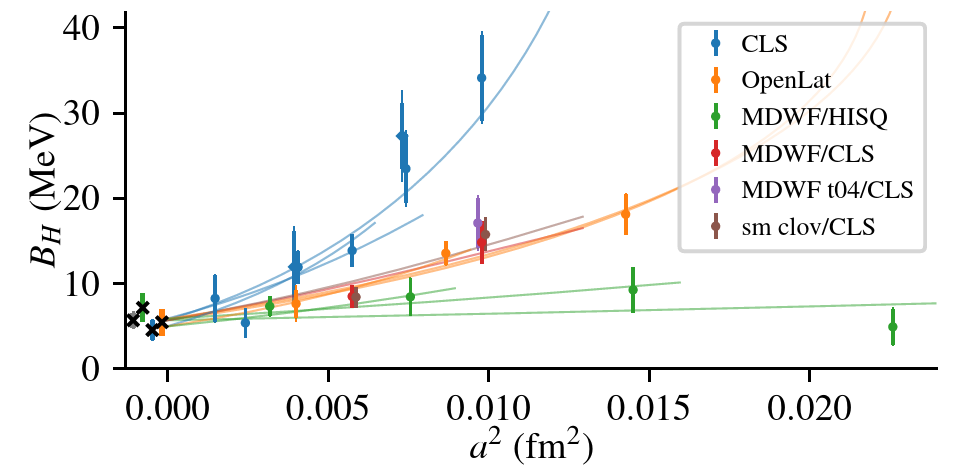}
\end{center}
\caption{The binding energy of the $H$-dibaryon at the $SU(3)_{\rm F}$ symmetric point
from Refs.~\cite{Green:2021qol,Green:2025rel}, showing the sensitivity to 
discretization effects.  (Left) Results for different lattice sizes. (Right)
Comparisons to other works.
\label{fig:Hdibaryon}}
\end{figure}

In Ref.~\cite{Green:2021qol}, the $H$-dibaryon at the $SU(3)_{\rm F}$ symmetric point
has recently been studied.  Sensitivity of the $H$-dibaryon binding energy to discretization
effects has been investigated and is shown in Fig.~\ref{fig:Hdibaryon}.  Further details 
about this study were presented at this conference and can be found in 
Ref.~\cite{Green:2025rel}.

\begin{figure}
\begin{center}
\raisebox{2mm}{\includegraphics[width=2.3in]{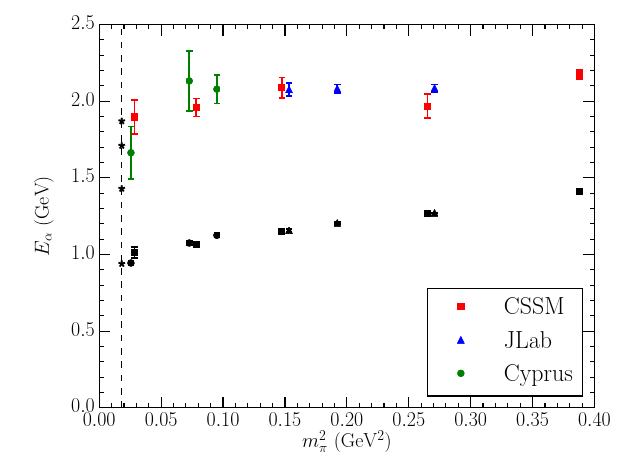}}\hspace*{1mm}
\includegraphics[width=3.6in]{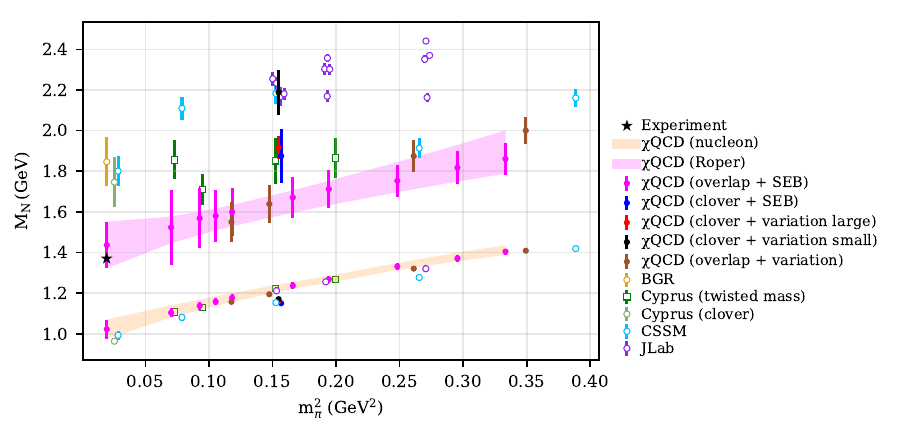}
\end{center}
\caption{(Left) Current status of positive-parity excitations of the proton
from Ref.~\cite{LeinweberNSTAR2024} which compares results from 
Refs.~\cite{Alexandrou:2014mka,Edwards:2011jj,Liu:2016uzk}.  Such studies
which only use three-quark operators miss the Roper and instead capture
a higher lying excitation. (Right) Results for the positive-parity
excitation spectrum of the proton from Ref.~\cite{xQCD:2019jke} are shown.
A large number of differently-smeared three-quark operators, combined with
a domain-wall fermion sea, overlap valence fermions, and a sequential Bayesian
analysis method, seems to capture the Roper (magenta band), but with rather
large uncertainties.
\label{fig:roper}}
\end{figure}

The first excitation of the proton, known as the Roper resonance, is an important
resonance.  Experimentally, it is a 4-star resonance $N(1440)$ with 
$I(J^P)=\frac{1}{2}(\frac{1}{2}^+)$ and a width in the range $250-450~{\rm MeV}$.
It is a notoriously difficult resonance to study in lattice QCD.  Three-quark
operators have difficulty capturing the Roper level near 1.4~GeV and instead yield 
an energy much higher near 2.0~GeV.  This fact is illustrated in the left hand plot
of Fig.~\ref{fig:roper} which shows energy extractions for the proton and its
first excitation from three lattice QCD studies.

Ref.~\cite{xQCD:2019jke} studied the Roper resonance using only a variety of 
three-quark operators with domain-wall fermions in the sea and overlap fermions
for the valence quarks.  Their results, shown on the right in Fig.~\ref{fig:roper},
are obtained using a large basis of three-quark operators with different smearings
and a sequential empirical Bayesian.  The Roper mass does seem to be captured,
but with very large uncertainties.

It has become clear that a definitive study of the Roper resonance needs 
multi-hadron operators involving $N\pi$, $N\sigma$, $\Delta\pi$ operators,
as well as $N\pi\pi$ operators.  Large volumes will be needed, as well as
a three-particle amplitude analysis, which is not yet available.

\section{Conclusion}

Novel lattice QCD methods, such as stochastic LapH and distillation, now
allow reliable determinations of energies involving multi-hadron states.
Large numbers of excited-state energy levels can be estimated, allowing
scattering phase shifts to be computed and hadron resonance properties,
such as masses and decay widths, to be determined.  In this talk,
recent results for the $\Delta$\ and $\Lambda(1405)$ resonances from
lattice QCD were highlighted, and the $NN$ discrepany at the $SU(3)_F$
symmetric point was discussed.  The good news is that this discrepancy
is now resolved and that current methods now seem to work well for
baryon-baryon scattering.  The Roper resonance is still a challenge,
but future studies involving three-particle operators may finally
shed light on this elusive hadron.  The author acknowledges support from 
the U.S.~NSF under awards PHY-2209167.

\bibliographystyle{JHEP}
\bibliography{references}

\providecommand{\href}[2]{#2}\begingroup\raggedright\begin{thebibliography}{10}

\bibitem{Metropolis:1953am}
N.~Metropolis, A.W.~Rosenbluth, M.N.~Rosenbluth, A.H.~Teller and E.~Teller,
  \emph{{Equation of state calculations by fast computing machines}},
  \href{https://doi.org/10.1063/1.1699114}{\emph{J. Chem. Phys.} {\bfseries 21}
  (1953) 1087}.

\bibitem{Clark:2006fx}
M.A.~Clark and A.D.~Kennedy, \emph{{Accelerating dynamical fermion computations
  using the rational hybrid Monte Carlo (RHMC) algorithm with multiple
  pseudofermion fields}},
  \href{https://doi.org/10.1103/PhysRevLett.98.051601}{\emph{Phys. Rev. Lett.}
  {\bfseries 98} (2007) 051601}
  [\href{https://arxiv.org/abs/hep-lat/0608015}{{\ttfamily hep-lat/0608015}}].

\bibitem{Morningstar:2003gk}
C.~Morningstar and M.J.~Peardon, \emph{{Analytic smearing of SU(3) link
  variables in lattice QCD}},
  \href{https://doi.org/10.1103/PhysRevD.69.054501}{\emph{Phys. Rev. D}
  {\bfseries 69} (2004) 054501}
  [\href{https://arxiv.org/abs/hep-lat/0311018}{{\ttfamily hep-lat/0311018}}].

\bibitem{HadronSpectrum:2009krc}
{\scshape Hadron Spectrum} collaboration, \emph{{A Novel quark-field creation
  operator construction for hadronic physics in lattice QCD}},
  \href{https://doi.org/10.1103/PhysRevD.80.054506}{\emph{Phys. Rev. D}
  {\bfseries 80} (2009) 054506}
  [\href{https://arxiv.org/abs/0905.2160}{{\ttfamily 0905.2160}}].

\bibitem{Morningstar:2011ka}
C.~Morningstar, J.~Bulava, J.~Foley, K.J.~Juge, D.~Lenkner, M.~Peardon et~al.,
  \emph{{Improved stochastic estimation of quark propagation with Laplacian
  Heaviside smearing in lattice QCD}},
  \href{https://doi.org/10.1103/PhysRevD.83.114505}{\emph{Phys. Rev. D}
  {\bfseries 83} (2011) 114505}
  [\href{https://arxiv.org/abs/1104.3870}{{\ttfamily 1104.3870}}].

\bibitem{Luscher:1990ux}
M.~Luscher, \emph{{Two particle states on a torus and their relation to the
  scattering matrix}},
  \href{https://doi.org/10.1016/0550-3213(91)90366-6}{\emph{Nucl. Phys. B}
  {\bfseries 354} (1991) 531}.

\bibitem{Luscher:1991cf}
M.~Luscher, \emph{{Signatures of unstable particles in finite volume}},
  \href{https://doi.org/10.1016/0550-3213(91)90584-K}{\emph{Nucl. Phys. B}
  {\bfseries 364} (1991) 237}.

\bibitem{Rummukainen:1995vs}
K.~Rummukainen and S.A.~Gottlieb, \emph{{Resonance scattering phase shifts on a
  nonrest frame lattice}},
  \href{https://doi.org/10.1016/0550-3213(95)00313-H}{\emph{Nucl. Phys. B}
  {\bfseries 450} (1995) 397}
  [\href{https://arxiv.org/abs/hep-lat/9503028}{{\ttfamily hep-lat/9503028}}].

\bibitem{Kim:2005gf}
C.h.~Kim, C.T.~Sachrajda and S.R.~Sharpe, \emph{{Finite-volume effects for
  two-hadron states in moving frames}},
  \href{https://doi.org/10.1016/j.nuclphysb.2005.08.029}{\emph{Nucl. Phys. B}
  {\bfseries 727} (2005) 218}
  [\href{https://arxiv.org/abs/hep-lat/0507006}{{\ttfamily hep-lat/0507006}}].

\bibitem{Briceno:2014oea}
R.A.~Briceno, \emph{{Two-particle multichannel systems in a finite volume with
  arbitrary spin}},
  \href{https://doi.org/10.1103/PhysRevD.89.074507}{\emph{Phys. Rev. D}
  {\bfseries 89} (2014) 074507}
  [\href{https://arxiv.org/abs/1401.3312}{{\ttfamily 1401.3312}}].

\bibitem{Morningstar:2017spu}
C.~Morningstar, J.~Bulava, B.~Singha, R.~Brett, J.~Fallica, A.~Hanlon et~al.,
  \emph{{Estimating the two-particle $K$-matrix for multiple partial waves and
  decay channels from finite-volume energies}},
  \href{https://doi.org/10.1016/j.nuclphysb.2017.09.014}{\emph{Nucl. Phys. B}
  {\bfseries 924} (2017) 477}
  [\href{https://arxiv.org/abs/1707.05817}{{\ttfamily 1707.05817}}].

\bibitem{Wigner:1946zz}
E.P.~Wigner, \emph{{Resonance Reactions and Anomalous Scattering}},
  \href{https://doi.org/10.1103/PhysRev.70.15}{\emph{Phys. Rev.} {\bfseries 70}
  (1946) 15}.

\bibitem{Wigner:1947zz}
E.P.~Wigner and L.~Eisenbud, \emph{{Higher Angular Momenta and Long Range
  Interaction in Resonance Reactions}},
  \href{https://doi.org/10.1103/PhysRev.72.29}{\emph{Phys. Rev.} {\bfseries 72}
  (1947) 29}.

\bibitem{Ross:1961jlg}
M.H.~Ross and G.L.~Shaw, \emph{{Multichannel effective range theory}},
  \href{https://doi.org/10.1016/0003-4916(61)90078-1}{\emph{Annals Phys.}
  {\bfseries 13} (1961) 147}.

\bibitem{deSwart:1962}
J.~{de Swart} and C.~Dullemond, \emph{Effective range theory and the low energy
  hyperon-nucleon interactions},
  \href{https://doi.org/https://doi.org/10.1016/0003-4916(62)90185-9}{\emph{Annals
  Phys.} {\bfseries 19} (1962) 458}.

\bibitem{Burke:2011}
P.~Burke, \emph{Chapter 3},  in \emph{{R-matrix theory of atomic collisions:
  Application to atomic, molecular and optical processes}}, (Heidelberg),
  pp.~135--139, Springer (2011).

\bibitem{Bulava:2022vpq}
J.~Bulava, A.D.~Hanlon, B.~H\"orz, C.~Morningstar, A.~Nicholson,
  F.~Romero-L\'opez et~al., \emph{{Elastic nucleon-pion scattering at
  $m_\pi=200$ MeV from lattice QCD}},
  \href{https://doi.org/10.1016/j.nuclphysb.2023.116105}{\emph{Nucl. Phys. B}
  {\bfseries 987} (2023) 116105}
  [\href{https://arxiv.org/abs/2208.03867}{{\ttfamily 2208.03867}}].

\bibitem{Alexandrou:2023elk}
C.~Alexandrou, S.~Bacchio, G.~Koutsou, T.~Leontiou, S.~Paul, M.~Petschlies
  et~al., \emph{{Elastic nucleon-pion scattering amplitudes in the
  \ensuremath{\Delta} channel at physical pion mass from lattice QCD}},
  \href{https://doi.org/10.1103/PhysRevD.109.034509}{\emph{Phys. Rev. D}
  {\bfseries 109} (2024) 034509}
  [\href{https://arxiv.org/abs/2307.12846}{{\ttfamily 2307.12846}}].

\bibitem{BaryonScatteringBaSc:2023zvt}
{\scshape Baryon Scattering (BaSc)} collaboration, \emph{{Two-Pole Nature of
  the \ensuremath{\Lambda}(1405) resonance from Lattice QCD}},
  \href{https://doi.org/10.1103/PhysRevLett.132.051901}{\emph{Phys. Rev. Lett.}
  {\bfseries 132} (2024) 051901}
  [\href{https://arxiv.org/abs/2307.10413}{{\ttfamily 2307.10413}}].

\bibitem{BaryonScatteringBaSc:2023ori}
{\scshape Baryon Scattering (BaSc)} collaboration, \emph{{Lattice QCD study of
  $\pi\Sigma-\overline{K}N$ scattering and the $Lambda(1405)$ resonance}},
  \href{https://doi.org/10.1103/PhysRevD.109.014511}{\emph{Phys. Rev. D}
  {\bfseries 109} (2024) 014511}
  [\href{https://arxiv.org/abs/2307.13471}{{\ttfamily 2307.13471}}].

\bibitem{Oller:2000fj}
J.A.~Oller and U.G.~Meissner, \emph{{Chiral dynamics in the presence of bound
  states: Kaon nucleon interactions revisited}},
  \href{https://doi.org/10.1016/S0370-2693(01)00078-8}{\emph{Phys. Lett. B}
  {\bfseries 500} (2001) 263}
  [\href{https://arxiv.org/abs/hep-ph/0011146}{{\ttfamily hep-ph/0011146}}].

\bibitem{Fukugita:1994ve}
M.~Fukugita, Y.~Kuramashi, M.~Okawa, H.~Mino and A.~Ukawa, \emph{{Hadron
  scattering lengths in lattice QCD}},
  \href{https://doi.org/10.1103/PhysRevD.52.3003}{\emph{Phys. Rev. D}
  {\bfseries 52} (1995) 3003}
  [\href{https://arxiv.org/abs/hep-lat/9501024}{{\ttfamily hep-lat/9501024}}].

\bibitem{NPLQCD:2012mex}
{\scshape NPLQCD} collaboration, \emph{{Light Nuclei and Hypernuclei from
  Quantum Chromodynamics in the Limit of SU(3) Flavor Symmetry}},
  \href{https://doi.org/10.1103/PhysRevD.87.034506}{\emph{Phys. Rev. D}
  {\bfseries 87} (2013) 034506}
  [\href{https://arxiv.org/abs/1206.5219}{{\ttfamily 1206.5219}}].

\bibitem{NPLQCD:2013bqy}
{\scshape NPLQCD} collaboration, \emph{{Nucleon-Nucleon Scattering Parameters
  in the Limit of SU(3) Flavor Symmetry}},
  \href{https://doi.org/10.1103/PhysRevC.88.024003}{\emph{Phys. Rev. C}
  {\bfseries 88} (2013) 024003}
  [\href{https://arxiv.org/abs/1301.5790}{{\ttfamily 1301.5790}}].

\bibitem{Berkowitz:2015eaa}
E.~Berkowitz, T.~Kurth, A.~Nicholson, B.~Joo, E.~Rinaldi, M.~Strother et~al.,
  \emph{{Two-Nucleon Higher Partial-Wave Scattering from Lattice QCD}},
  \href{https://doi.org/10.1016/j.physletb.2016.12.024}{\emph{Phys. Lett. B}
  {\bfseries 765} (2017) 285}
  [\href{https://arxiv.org/abs/1508.00886}{{\ttfamily 1508.00886}}].

\bibitem{Inoue:2011ai}
{\scshape HAL QCD} collaboration, \emph{{Two-Baryon Potentials and H-Dibaryon
  from 3-flavor Lattice QCD Simulations}},
  \href{https://doi.org/10.1016/j.nuclphysa.2012.02.008}{\emph{Nucl. Phys. A}
  {\bfseries 881} (2012) 28} [\href{https://arxiv.org/abs/1112.5926}{{\ttfamily
  1112.5926}}].

\bibitem{Horz:2020zvv}
B.~H\"orz et~al., \emph{{Two-nucleon S-wave interactions at the $SU(3)$
  flavor-symmetric point with $m_{ud}\simeq m_s^{\rm phys}$: A first lattice
  QCD calculation with the stochastic Laplacian Heaviside method}},
  \href{https://doi.org/10.1103/PhysRevC.103.014003}{\emph{Phys. Rev. C}
  {\bfseries 103} (2021) 014003}
  [\href{https://arxiv.org/abs/2009.11825}{{\ttfamily 2009.11825}}].

\bibitem{Francis:2018qch}
A.~Francis, J.R.~Green, P.M.~Junnarkar, C.~Miao, T.D.~Rae and H.~Wittig,
  \emph{{Lattice QCD study of the $H$ dibaryon using hexaquark and two-baryon
  interpolators}},
  \href{https://doi.org/10.1103/PhysRevD.99.074505}{\emph{Phys. Rev. D}
  {\bfseries 99} (2019) 074505}
  [\href{https://arxiv.org/abs/1805.03966}{{\ttfamily 1805.03966}}].

\bibitem{Amarasinghe:2021lqa}
S.~Amarasinghe, R.~Baghdadi, Z.~Davoudi, W.~Detmold, M.~Illa, A.~Parreno
  et~al., \emph{{Variational study of two-nucleon systems with lattice QCD}},
  \href{https://doi.org/10.1103/PhysRevD.107.094508}{\emph{Phys. Rev. D}
  {\bfseries 107} (2023) 094508}
  [\href{https://arxiv.org/abs/2108.10835}{{\ttfamily 2108.10835}}].

\bibitem{NNsu3}
J.~Bulava, M.~Clark, A.S.~Gambhir, A.D.~Hanlon, B.~H\"orz, B.~Jo\'o et~al.,
  \emph{{Di-nucleons do not form bound states at heavy pion mass}},
  \href{https://arxiv.org/abs/2505.05547}{{\ttfamily 2505.05547}}.

\bibitem{Green:2021qol}
J.R.~Green, A.D.~Hanlon, P.M.~Junnarkar and H.~Wittig, \emph{{Weakly bound $H$
  dibaryon from SU(3)-flavor-symmetric QCD}},
  \href{https://doi.org/10.1103/PhysRevLett.127.242003}{\emph{Phys. Rev. Lett.}
  {\bfseries 127} (2021) 242003}
  [\href{https://arxiv.org/abs/2103.01054}{{\ttfamily 2103.01054}}].

\bibitem{Green:2022rjj}
{\scshape Baryon Scattering (BaSc)} collaboration, \emph{{Nucleon-nucleon
  scattering from distillation}},
  \href{https://doi.org/10.22323/1.430.0200}{\emph{PoS} {\bfseries LATTICE2022}
  (2023) 200} [\href{https://arxiv.org/abs/2212.09587}{{\ttfamily
  2212.09587}}].

\bibitem{Geng:2024dpk}
Y.~Geng, L.~Liu, P.~Sun, J.-J.~Wu, H.~Xing, Z.~Yan et~al., \emph{{Doubly
  Charmed $H$-like dibaryon $\Lambda_c \Lambda_c$ scattering from Lattice
  QCD}}, \href{https://doi.org/10.22323/1.466.0307}{\emph{PoS} {\bfseries
  LATTICE2024} (2025) 307}.

\bibitem{Beane:2017edf}
S.R.~Beane et~al., \emph{{Comment on ``Are two nucleons bound in lattice QCD
  for heavy quark masses? - Sanity check with L\"uscher's finite volume formula
  -''}},  \href{https://arxiv.org/abs/1705.09239}{{\ttfamily 1705.09239}}.

\bibitem{Iritani:2018vfn}
{\scshape HAL QCD} collaboration, \emph{{Consistency between
  L\"uscher\textquoteright{}s finite volume method and HAL QCD method for
  two-baryon systems in lattice QCD}},
  \href{https://doi.org/10.1007/JHEP03(2019)007}{\emph{JHEP} {\bfseries 03}
  (2019) 007} [\href{https://arxiv.org/abs/1812.08539}{{\ttfamily
  1812.08539}}].

\bibitem{Green:2025rel}
J.R.~Green, \emph{{Status of two-baryon scattering in lattice QCD}},  in
  \emph{{11th International Workshop on Chiral Dynamics}}, 2, 2025
  [\href{https://arxiv.org/abs/2502.15546}{{\ttfamily 2502.15546}}].

\bibitem{xQCD:2019jke}
{\scshape xQCD} collaboration, \emph{{Roper State from Overlap Fermions}},
  \href{https://doi.org/10.1103/PhysRevD.101.054511}{\emph{Phys. Rev. D}
  {\bfseries 101} (2020) 054511}
  [\href{https://arxiv.org/abs/1911.02635}{{\ttfamily 1911.02635}}].

\bibitem{LeinweberNSTAR2024}
D.~Leinweber, \emph{{Physical Interpretation of the Baryon Spectrum}},  in
  \emph{{14th International Workshop on the Physics of Excited Nucleons}},
  2024.

\bibitem{Alexandrou:2014mka}
C.~Alexandrou, T.~Leontiou, C.N.~Papanicolas and E.~Stiliaris, \emph{{Novel
  analysis method for excited states in lattice QCD: The nucleon case}},
  \href{https://doi.org/10.1103/PhysRevD.91.014506}{\emph{Phys. Rev. D}
  {\bfseries 91} (2015) 014506}
  [\href{https://arxiv.org/abs/1411.6765}{{\ttfamily 1411.6765}}].

\bibitem{Edwards:2011jj}
R.G.~Edwards, J.J.~Dudek, D.G.~Richards and S.J.~Wallace, \emph{{Excited state
  baryon spectroscopy from lattice QCD}},
  \href{https://doi.org/10.1103/PhysRevD.84.074508}{\emph{Phys. Rev. D}
  {\bfseries 84} (2011) 074508}
  [\href{https://arxiv.org/abs/1104.5152}{{\ttfamily 1104.5152}}].

\bibitem{Liu:2016uzk}
Z.-W.~Liu, W.~Kamleh, D.B.~Leinweber, F.M.~Stokes, A.W.~Thomas and J.-J.~Wu,
  \emph{{Hamiltonian effective field theory study of the $N^*(1440)$ resonance
  in lattice QCD}},
  \href{https://doi.org/10.1103/PhysRevD.95.034034}{\emph{Phys. Rev. D}
  {\bfseries 95} (2017) 034034}
  [\href{https://arxiv.org/abs/1607.04536}{{\ttfamily 1607.04536}}].

\end{thebibliography}\endgroup

\end{document}